\documentclass[prl,twocolumn,lengthcheck,superscriptaddress,letterpaper,nofootinbib,nopreprintnumbers,showpacs]{revtex4-1}

\usepackage{color}
\usepackage{graphicx}
\usepackage{float}
\usepackage{amsmath}
\usepackage{acronym}
\usepackage{multirow}
\usepackage{lineno}
\usepackage{svn-multi}
\usepackage{import}
\usepackage{hyperref}
\usepackage[normalem]{ulem}
\usepackage[caption=false]{subfig}
\usepackage{pifont}
\usepackage{etoolbox}

\renewcommand{\today}{\number\day\space\ifcase\month\or
  January\or February\or March\or April\or May\or June\or
  July\or August\or September\or October\or November\or December\fi
  \space\number\year}

\newcommand{\macro}[1]{#1} 

\newcommand{\Msun}{\ensuremath{\mathrm{M}_\odot}}

\newcommand{\RadiusheavyOR}{\macro{\ensuremath{11.8_{-3.3}^{+2.7}}}} 
\newcommand{\RadiuslightOR}{\macro{\ensuremath{10.8_{-3.0}^{+2.9}}}} 
\newcommand{\DeltaRlightOR}{\macro{\ensuremath{5.9}}} 

\newcommand{\ERRORCofLambda}{\macro{\ensuremath{6.5}}} 
\newcommand{\SIGMACofLambda}{\macro{\ensuremath{3}}} 

\newcommand{\Lambdameasurement}{\macro{\ensuremath{190^{+390}_{-120}}}} 

\newcommand{\RATIOCAfromUR}{\macro{\ensuremath{3}}} 

\newcommand{\RadiusheavyUR}{\macro{\ensuremath{10.8_{-1.7}^{+2.0}}}} 
\newcommand{\RadiuslightUR}{\macro{\ensuremath{10.7_{-1.5}^{+2.1}}}} 
\newcommand{\DeltaRlightUR}{\macro{\ensuremath{3.6}}} 

\newcommand{\massoneUR}{\macro{\ensuremath{(1.15,1.36)}}} 
\newcommand{\masstwoUR}{\macro{\ensuremath{(1.36,1.62)}}} 

\newcommand{\Pattwicesaturation}{\macro{\ensuremath{3.5_{-1.7}^{+2.7}}}} 
\newcommand{\Patsixtimessaturation}{\macro{\ensuremath{9.0_{-2.6}^{+7.9}}}} 

\newcommand{\RadiusheavyPPmm}{\macro{\ensuremath{11.9_{-1.4}^{+1.4}}}} 
\newcommand{\RadiuslightPPmm}{\macro{\ensuremath{11.9_{-1.4}^{+1.4}}}} 
\newcommand{\DeltaRlightPPmm}{\macro{\ensuremath{2.8}}} 

\newcommand{\Radiuscutoof}{\macro{\ensuremath{10}}} 

\newcommand{\Maxmass}{\macro{\ensuremath{1.97}}} 

\newcommand{\massoneS}{\macro{\ensuremath{(1.18,1.36)}}} 
\newcommand{\masstwoS}{\macro{\ensuremath{(1.36,1.58)}}} 

\graphicspath{{figures/}} 

\usepackage{draftwatermark}

\SetWatermarkAngle{90}
\SetWatermarkLightness{1.0}
\SetWatermarkFontSize{0.5cm}
\SetWatermarkScale{1}
\SetWatermarkText{draft}


\acrodef{BBH}[BBH]{binary black hole}
\acrodef{BNS}[BNS]{binary neutron star}
\acrodef{NS}[NS]{neutron star}
\acrodef{EOS}[EOS]{equation of state}
\acrodef{BHNS}[BHNS]{black hole--neutron star binaries}
\acrodef{PBH}[PBH]{primordial black hole binaries}
\acrodef{SNR}[SNR]{signal-to-noise ratio}
\acrodef{LIGO}[LIGO]{Laser Interferometer Gravitational-wave Observatory}
\acrodef{LHO}[LHO]{LIGO Hanford}
\acrodef{LLO}[LLO]{LIGO Livingston}
\acrodef{LSC}[LSC]{LIGO Scientific Collaboration}
\acrodef{CBC}[CBC]{compact binary coalescence}
\acrodef{GW}[GW]{gravitational wave}
\acrodef{GW}[GW]{gravitational wave}
\acrodef{FAR}[FAR]{false alarm rate}
\acrodef{FAP}[FAP]{false alarm probability}
\acrodef{IFO}[IFO]{interferometer}
\acrodef{BH}[BH]{black hole}
\acrodef{NR}[NR]{numerical relativity}
\acrodef{PN}[PN]{post-Newtonian}
\acrodef{EOB}[EOB]{effective-one-body}
\acrodef{PSD}[PSD]{power spectral density}
\acrodef{PDF}[PDF]{probability density function}
\acrodef{PE}[PE]{parameter-estimation}
\acrodef{LAL}[LAL]{LIGO Algorithm Library}

\newcommand{\event}{GW170817}
\newcommand{\tf}{TaylorF2}

\newcommand{\imrpp}{PhenomPNRT}

\newtoggle{endauthorlist}
\toggletrue{endauthorlist}

\begin{document}

\title{GW170817: Measurements of Neutron Star Radii and Equation of State}

\iftoggle{endauthorlist}{
  %
  %
  \let\mymaketitle\maketitle
  \let\myauthor\author
  \let\myaffiliation\affiliation
  \author{The LIGO Scientific Collaboration}
  \author{The Virgo Collaboration}
}{

\author{B.~P.~Abbott}
\affiliation{LIGO, California Institute of Technology, Pasadena, CA 91125, USA}
\author{R.~Abbott}
\affiliation{LIGO, California Institute of Technology, Pasadena, CA 91125, USA}
\author{T.~D.~Abbott}
\affiliation{Louisiana State University, Baton Rouge, LA 70803, USA}
\author{F.~Acernese}
\affiliation{Universit\`a di Salerno, Fisciano, I-84084 Salerno, Italy}
\affiliation{INFN, Sezione di Napoli, Complesso Universitario di Monte S.Angelo, I-80126 Napoli, Italy}
\author{K.~Ackley}
\affiliation{OzGrav, School of Physics \& Astronomy, Monash University, Clayton 3800, Victoria, Australia}
\author{C.~Adams}
\affiliation{LIGO Livingston Observatory, Livingston, LA 70754, USA}
\author{T.~Adams}
\affiliation{Laboratoire d'Annecy de Physique des Particules (LAPP), Univ. Grenoble Alpes, Universit\'e Savoie Mont Blanc, CNRS/IN2P3, F-74941 Annecy, France}
\author{P.~Addesso}
\affiliation{University of Sannio at Benevento, I-82100 Benevento, Italy and INFN, Sezione di Napoli, I-80100 Napoli, Italy}
\author{R.~X.~Adhikari}
\affiliation{LIGO, California Institute of Technology, Pasadena, CA 91125, USA}
\author{V.~B.~Adya}
\affiliation{Max Planck Institute for Gravitational Physics (Albert Einstein Institute), D-30167 Hannover, Germany}
\affiliation{Leibniz Universit\"at Hannover, D-30167 Hannover, Germany}
% ...

\collaboration{The LIGO Scientific Collaboration and the Virgo Collaboration}

  \author{The LIGO Scientific Collaboration}
  \affiliation{The LSC}
  \author{The Virgo Collaboration}
  \affiliation{Virgo}
}

\begin{abstract}
On 17 August 2017, the LIGO and Virgo observatories made the first direct detection of gravitational waves from the coalescence of a neutron star binary system.
The detection of this gravitational-wave signal, \event{}, offers a novel opportunity to directly probe the properties of matter at the extreme conditions found in the interior of these stars.
The initial, minimal-assumption analysis of the LIGO and Virgo data placed constraints on the tidal effects of the coalescing bodies, which were then translated to constraints on neutron star radii.
Here, we expand upon previous analyses by working under the hypothesis that both bodies were neutron stars that are described by the same equation of state and have spins within the range observed in Galactic binary neutron stars.
Our analysis employs two methods: the use of equation-of-state-insensitive relations between various macroscopic properties of the neutron stars and the use of an efficient parametrization of the defining function $p(\rho)$ of the equation of state itself. 
From the LIGO and Virgo data alone and the first method, we measure the two neutron star radii as $R_1= \RadiusheavyUR$\,km for the heavier star and $R_2= \RadiuslightUR$\,km for the lighter star at the 90\% credible level.
If we additionally require that the equation of state supports neutron stars with masses larger than $\Maxmass \,\Msun$ as required from electromagnetic observations and employ the equation-of-state parametrization, we further constrain $R_1= \RadiusheavyPPmm$\,km and $R_2= \RadiuslightPPmm$\,km at the 90\% credible level.
Finally, we obtain constraints on $p(\rho)$ at supranuclear densities, with 
pressure at twice nuclear saturation density measured at
$\Pattwicesaturation \times 10^{34}\,\mathrm{dyn}\,\mathrm{cm}^{-2}$ 
at the 90\% level.
\end{abstract}

\maketitle

\section{Introduction}

Since September 2015, the Advanced LIGO~\cite{cqg.32.074001.15} and Advanced
Virgo~\cite{TheVirgo:2014hva} observatories have opened a window
on the gravitational-wave (GW) universe~\cite{GW150914_paper,Aasi:2013wya}.
A new type of astrophysical source of GWs was detected on 17 August 2017,
when the GW signal emitted by a low-mass coalescing compact binary was
observed~\cite{DetectionPaper}.
This observation coincided with the detection of a $\gamma$-ray burst, GRB 170817A~\cite{Monitor:2017mdv,Goldstein:2017mmi},
verifying that the source binary contained matter, which was further corroborated
by a series of observations that followed across the
electromagnetic spectrum; see e.g.~\cite{GBM:2017lvd,Coulter:2017wya,Troja:2017nqp,Haggard:2017qne,Hallinan:2017woc}.
The measured masses of the bodies and the variety of electromagnetic observations
are consistent with neutron stars (NSs).

Neutron stars are unique natural laboratories for studying the behavior of cold
high-density nuclear matter.
Such behavior is governed by the equation of state (EOS), which prescribes a relationship between
pressure and density.
This determines the relation between NS mass and
radius, as well as other macroscopic properties such as the stellar moment of inertia and
the tidal deformability (see e.g.~\cite{2015PhRvC..91a5804S}).
While terrestrial experiments are able to test and constrain the cold EOS at
densities below and near the saturation density of nuclei $\rho_\mathrm{nuc}=2.8\times 10^{14}$\,g cm$^{-3}$ (see e.g.
\cite{2012PhRvC..86a5803T,2016PrPNP..91..203B,2016PhR...621..127L,2017RvMP...89a5007O}
for a review), currently they cannot probe
the extreme conditions in the deep core of NSs.
Astrophysical measurements of NS masses, radii, moments of inertia and tidal effects, on the other hand, have the potential 
to offer information about whether the EOS is soft or stiff and what the pressure is at several
times the nuclear saturation density
\cite{2013ApJ...773...11H,2016PhR...621..127L,Ozel:2016oaf,Steiner:2017vmg}.

GWs offer an opportunity for such astrophysical measurements to be performed, as the GW
signal emitted by merging NS binaries differs from that of two merging black holes (BHs).
The most prominent effect of matter during the observed binary inspiral comes from the tidal deformation
that each star's gravitational field induces on its companion. This
deformation enhances GW emission and thus accelerates
 the decay of the quasicircular inspiral~\cite{flanagan:021502,Vines:2011ud,PhysRevD.85.123007}.
In the post-Newtonian (PN) expansion of the inspiral dynamics~\cite{Buonanno:2009zt,Blanchet:2013haa,Goldberger:2004jt,Goldberger:2007hy,Damour:2000zb,Blanchet:1995ez,Blanchet:2004ek,Damour:2001bu,Flanagan:1997fn}, 
this effect causes the phase of the GW signal to differ from that of a binary BH (BBH) from the fifth PN order onwards~\cite{Damour:1982wm,flanagan:021502,Gralla:2017djj}. 
The leading-order contribution is proportional to each star's tidal deformability parameter,
$\Lambda = (2/3) k_{2} C^{-5}$, an EOS-sensitive quantity that describes how much a star is deformed in the presence of a tidal field. Here $k_2$ is the $l=2$ relativistic Love number~\cite{Damour:1983a,Hinderer:2007mb,Hinderer:2009ca,Binnington:2009bb,Damour:2009vw}, 
$C\equiv G m/(c^2 R)$ is the compactness, $R$ is the areal radius, and $m$ is the mass of the NS.
The deformation of each NS due to its own spin also modifies
the waveform and depends on the EOS. 
This effect enters the post-Newtonian expansion as a contribution to the (lowest order) spin-spin term at the second order in the GW phase~\cite{Poisson:1997ha,Bohe:2015ana}.
The EOS also affects the waveform at merger, the merger outcome and its lifetime, as well as the postmerger emission (see e.g.~\cite{Baiotti:2016qnr}).
Finally, other stellar modes can couple to the tidal field and affect the GW signal~\cite{1994MNRAS.270..611L,flanagan:021502,2016PhRvL.116r1101H,2018PhRvD..97b3016A}.

Among the various EOS-dependent effects, the tidal deformation is the one most readily measurable with \event{}.
The spin-induced quadrupole has a larger effect on the orbital evolution for systems
with large NS spin
~\cite{Laarakkers:1997hb,Pappas:2012qg,PhysRevD.92.023012,Harry:2018hke} but is also largely degenerate with the mass ratio and
 the NS spins, making it difficult to measure independently~\cite{Krishnendu:2017shb}. The postmerger
signal, while rich in content, is also difficult to observe, with current detector sensitivities 
being limited due to photon shot noise~\cite{cqg.32.074001.15} at the high frequencies of interest. The merger and postmerger 
signal make a negligible contribution to our inference for GW170817~\cite{PostmergerSearchPaper,Abbott:2018wiz}.

In~\cite{DetectionPaper}, we presented the first measurements of the properties of
GW170817, including a first set of constraints on the tidal deformabilities of
the two compact objects, from which inferences about the EOS can be made.
An independent analysis further exploring how well the gravitational-wave data can be used
to constrain the tidal deformabilites, and, from that, the NS radii, has
also been performed recently~\cite{De:2018uhw}.
Our initial bounds have facilitated a large number of studies,
e.g.~\cite{2041-8205-850-2-L19,2041-8205-850-2-L34,Zhou:2017pha,2041-8205-852-2-L25,
PhysRevLett.120.172702,Nandi:2017rhy,Paschalidis:2017qmb,PhysRevD.97.021501,
PhysRevLett.120.172703,Raithel:2018ncd,Most:2018hfd}, aiming to translate the measurements
of masses and tidal deformabilities into constraints on the EOS of NS matter. 
In a companion paper \cite{Abbott:2018wiz}, we perform a more detailed analysis focusing on the source properties, 
improving upon the original analysis of~\cite{DetectionPaper} by using Virgo data with reduced
calibration uncertainty,
extending the analysis to lower frequencies, employing more accurate waveform models, and fixing
the location of the source in the sky to the one identified by the electromagnetic observations.

Here we complement the analysis of~\cite{Abbott:2018wiz}, 
and work under the hypothesis that GW170817 was the result of a
coalescence of two NSs whose masses and spins
are consistent with astrophysical observations and expectations. Moreover since
NSs represent equilibrium ground-state configurations, we assume that
their properties are described by the same EOS.
By making these additional assumptions, we are able to further improve our measurements
of the tidal deformabilities of GW170817, and constrain the radii of the two NSs.
Moreover, we use an efficient parametrization of the EOS to place constraints on the 
pressure of cold matter at supranuclear densities using GW observations. 
This direct measurement of the pressure takes into account physical and observational
constraints on the NS EOS, namely causality, 
thermodynamic stability, and
a lower limit on the maximum NS mass supported by the EOS to be $M_{\textrm{max}}>\Maxmass\,\Msun$.
The latter is chosen as a $1\sigma$ conservative estimate, based on the observation of PSR J0348+0432 with $M=2.01\pm 0.04\,\Msun$\cite{Antoniadis:2013pzd}, the heaviest NS known to date.

The radii measurements presented here improve upon existing results (e.g.~\cite{PhysRevLett.120.172703,PhysRevLett.120.172702}) which had used the initial tidal measurements reported in~\cite{DetectionPaper}.
We also verify that our radii measurements are consistent with the result of the methodologies presented in these studies when applied to our improved tidal measurements.
Moreover, we obtain a more precise estimate of the NS radius than~\cite{De:2018uhw}.

\section{Methods}
\label{sec:method}

In this section we describe the details of the analysis. We use the same LIGO and Virgo data and calibration model
analyzed in~\cite{Abbott:2018wiz}. The data can be downloaded from the Gravitational Wave Open Science Center (GWOSC)~\cite{GWOSC}.
The data include the subtraction of an instrumental artifact occurring at LIGO-Livingston within 2~s of the GW170817 merger~\cite{DetectionPaper,Pankow:2018qpo},
as well as the subtraction of independently measurable noise sources~\cite{Driggers:2018gii,iLIGO_SeisCleaning,iLIGO_LSCCleaning,iLIGO_PEMCleaning}.

\subsection{Bayesian methods}
\label{Sec:Bayesian}

We employ a coherent Bayesian analysis to estimate the source parameters $\vec{\vartheta}$
as described in~\cite{Veitch:2014wba,TheLIGOScientific:2016wfe}. The goal is to determine 
the posterior probability density function (PDF), $p(\vec{\vartheta} | d)$, given
the LIGO and Virgo data $d$. Given a prior PDF $p(\vec{\vartheta})$ on the parameter space (quantifying our prior
belief in observing a source with properties $\vec{\vartheta}$), the posterior PDF is given by Bayes's theorem
$p(\vec{\vartheta} | d) \propto p(\vec{\vartheta}) p(d | \vec{\vartheta})$, where $p(d | \vec{\vartheta})$ is the likelihood 
 of obtaining the data $d$ given that a signal with parameters $\vec{\vartheta}$ is present in the data.
Evaluating the multidimensional $p(\vec{\vartheta} | d)$ 
analytically is computationally prohibitive
so we resort to sampling techniques to efficiently
draw samples from the underlying distribution.
We make use of the Markov-chain Monte Carlo algorithm as implemented in the \textsc{LALInference} package~\cite{Veitch:2014wba},
which is part of the publicly available LSC Algorithm Library (\textsc{LAL})~\cite{lalinference_o2}.
For the likelihood calculation, we use $128$~s of data around GW170817 over a frequency range of 
23--2048~Hz, covering both the time and frequency ranges where there was appreciable signal above 
the detector noise, and we estimate the likelihood for our waveform templates up to merger.
The power spectral density (PSD) of the noise is computed on source~\cite{Littenberg:2014oda,CoLi2015,Abbott:2018wiz},
and we marginalize over the detectors' calibration uncertainties as described in~\cite{SplineCalMarg-T1400682,Abbott:2018wiz,TheLIGOScientific:2016wfe}.

In the analysis of a GW signal from a binary NS coalescence, 
the source parameters $\vec{\vartheta}$ on which the signal depends can be decomposed as $\vec{\vartheta} =  (\vec{\vartheta}_{\mathrm{PM}}, \vec{\vartheta}_{\mathrm{EOS}})$, into parameters
that would be present if the two bodies behaved like point masses $\vec{\vartheta}_{\mathrm{PM}}$, and EOS-sensitive parameters $\vec{\vartheta}_{\mathrm{EOS}}$ that arise due
to matter effects of the two finite-sized bodies (e.g. tidal deformabilities).
The priors on the point-mass parameters that we use are described in Sec.~II~D of~\cite{Abbott:2018wiz} and we do not
repeat them here. 
We also use the same convention for the component masses, i.e. $m_1 \geq m_2$.
We only consider the ``low-spin'' prior of~\cite{Abbott:2018wiz} where the dimensionless NS spin parameter is restricted to $\chi \leq \, 0.05$, in agreement with expectations from Galactic binary NS spin measurements~\cite{Tauris:2017omb}, and we fix the location of the source in the sky to the one given by EM observations.
Regarding the EOS-related part of the parameter space and the corresponding priors, we consider two physically motivated parameterizations of different dimensionalities, which we describe in detail in the following sections.
The first method requires the sampling of tidal deformability parameters, whereas the second method directly samples the EOS function $p(\rho)$ from a 4-dimensional family of functions.
In both cases, the assumption that the binary consists of two NSs that are described by the same EOS is implicit in the parametrization of matter effects (in contrast with the analysis of~\cite{Abbott:2018wiz}, where minimal assumptions are made about the nature of the source).

\subsection{Waveform models and matter effects}
The measurement process described above 
requires a waveform model that maps the source parameters $\vec{\vartheta}$ to
a signal $h(t;\vec{\vartheta})$ that would be observed in the detector. The publicly available
\textsc{LALSimulation} software package of \textsc{LAL}~\cite{lalinference_o2} contains several
such waveform models obtained with different theoretical approaches. 
The impact of varying the models among several choices~\cite{Schmidt:2012rh,Hannam:2013oca,Schmidt:2014iyl,Husa:2015iqa,Khan:2015jqa,Dietrich:2017aum,Bohe:2016gbl,Bernuzzi:2014owa,Lackey:2016krb,Hinderer:2016eia,Steinhoff:2016rfi,Sathyaprakash:1991mt,Bohe:2013cla, Arun:2008kb, Mikoczi:2005dn, Bohe:2015ana, Mishra:2016whh,flanagan:021502,Vines:2011ud} is analyzed in detail in~\cite{Abbott:2018wiz}, showing that for GW170817 the systematic uncertainties due to the modeling of matter effects or the underlying point-particle description are smaller than the statistical errors in the measurement. We perform a similar analysis here by varying the BBH baseline model or using a post-Newtonian waveform prescription and find results consistent with those presented in Sec.~III~D and Table IV of~\cite{Abbott:2018wiz}. 
Moreover, the study of Appendix A of~\cite{Abbott:2018wiz} (Table V) suggests that varying the tidal description in the waveforms also leads to broadly consistent tidal measurements.
Since the net effect of varying waveform models is very different for each of the source properties, we refer to the tables and figures in~\cite{Abbott:2018wiz} for quantitative statements to assess the impact of modeling uncertainties. 

In the \event{} discovery paper~\cite{DetectionPaper} the results for the inferred tidal deformabilities were obtained with the \tf{} model that is based solely on post-Newtonian results for both the BBH baseline model~\cite{Sathyaprakash:1991mt,Bohe:2013cla, Arun:2008kb, Mikoczi:2005dn, Bohe:2015ana, Mishra:2016whh} and for tidal effects~\cite{flanagan:021502,Vines:2011ud}, as this model led to the conservatively largest bounds. 
In this Letter, we use a more realistic waveform model \imrpp{}~\cite{Schmidt:2012rh,Hannam:2013oca,Schmidt:2014iyl,Husa:2015iqa,Khan:2015jqa}, which is also used as the reference model in our detailed analysis of the properties of \event{}~\cite{Abbott:2018wiz}. The BBH baseline in this model, constructed based on~\cite{Taracchini:2013,Bohe:2013cla, Buonanno:2009zt,Arun:2008kb,Mikoczi:2005dn,Barausse:2009xi, Buonanno:2000ef, BuonannoDamour:1999}, is calibrated to numerical relativity data and describes relativistic point-mass, spin, and the dominant precession effects. The model further includes tidal effects in the phase from combining analytical information~\cite{Bernuzzi:2014owa, Wade:2014vqa,PhysRevD.85.123007,Vines:2011ud} with results from numerical-relativity simulations of binary NSs as described in~\cite{Dietrich:2017aum, Dietrich:2018uni}, and matter effects in the spin-induced quadrupole based post-Newtonian results~\cite{Poisson:1997ha,Bohe:2015ana,Arun:2008kb,Mikoczi:2005dn,Mishra:2016whh}. The characteristic rotational quadrupole deformation parameters are computed from $\Lambda$ through EOS-insensitive relations~\cite{Yagi:2013bca,Yagi:2013awa} as described in ~\cite{PhysRevD.92.023012,Chatziioannou:2015uea}. Other matter effects with nonzero spins are not taken into account in our analysis.

\subsection{EOS-insensitive relations}

Despite the microscopic complexity of NSs, some of their macroscopic properties are linked
by EOS-insensitive relations that depend only weakly on the EOS~\cite{Yagi:2016bkt}. 
We use two such relations to ensure that the two NSs obey the same EOS and to translate
NS tidal deformabilities to NS radii. 

The first such relation we employ was constructed in~\cite{Yagi:2015pkc}
and studied in the context of realistic GW inference in~\cite{Chatziioannou:2018vzf}. 
It combines the mass ratio of the binary $q\equiv m_2/m_1 \leq 1$, the symmetric tidal deformability
$\Lambda_s \equiv (\Lambda_2+\Lambda_1)/2$ and the antisymmetric tidal deformability
$\Lambda_a \equiv (\Lambda_2-\Lambda_1)/2$ in a relation of the form $\Lambda_a(\Lambda_s,q)$. 
Fitting coefficients and an estimate of the relation's intrinsic error were obtained by tuning
to a large set of EOS models~\cite{Yagi:2016bkt,Chatziioannou:2018vzf}, ensuring that the relation gives pairs of tidal deformabilities
that correspond to realistic EOS models.
We sample uniformly in the symmetric tidal deformability $\Lambda_s \in [0,5000]$,
use the EOS-insensitive relation to compute $\Lambda_a$, and then obtain $\Lambda_1$ and $\Lambda_2$,
which are used to generate a waveform template.
The sampling of tidal parameters also involves a marginalization over the intrinsic error in the relation, which is also a function of $\Lambda_s$ and $q$.
This procedure leads to unbiased estimation of the tidal parameters for a wide range of EOSs and mass ratios~\cite{Chatziioannou:2018vzf}.

The second relation we employ is between NS tidal deformability $\Lambda$ and NS compactness $C$~\cite{Maselli:2013mva,Urbanec:2013fs}.
We employ this $\Lambda$--$C$ relation with the coefficients given in Sec.~(4.4) of~\cite{Yagi:2016bkt} to compute the posterior for the radius and the mass of each binary component. Reference~\cite{Yagi:2016bkt} reports a maximum $\ERRORCofLambda\%$ relative error in the relation when compared to a large set of EOS models.
We assume that the relative error is constant across the parameter space and distributed according to a zero-mean Gaussian with a standard deviation of $(\ERRORCofLambda/\SIGMACofLambda)\%$ and marginalize over it.
We verified that our results are not sensitive to this choice of error estimate by comparing to the more conservative choice of a uniform distribution in $[-6.5\%,6.5\%]$.

\subsection{Parametrized EOS}

Instead of sampling macroscopic EOS-related parameters such as tidal deformabilities, one may instead sample the defining function $p(\rho)$ of the EOS directly.
A number of parametrizations of different degrees of complexity and fidelity to
realistic EOS models have been proposed (see~\cite{Lindblom:2018rfr} for a review), and here we employ the spectral parametrization
constructed and validated in~\cite{PhysRevD.82.103011,2012PhRvD..86h4003L,Lindblom:2013kra}.
This parametrization expresses the logarithm of the adiabatic index of the EOS $\Gamma(p;\gamma_i)$, as a polynomial of the
pressure $p$, where $\gamma_i=(\gamma_0,\gamma_1,\gamma_2,\gamma_3)$
are the free EOS parameters. The adiabatic index is then used to compute the energy
$\epsilon(p;\gamma_i)$ and rest-mass density $\rho(p;\gamma_i)$, which are inverted to give the EOS.
The parameterized high-density EOS is then stitched to the SLy EOS~\cite{Douchin:2001sv} below about half
the nuclear saturation density.
This is chosen because such low densities do not significantly impact the global properties of the NS~\cite{Lattimer:2000nx}.
Different low density EOSs can produce a difference in radius, for a given $m$, of order 0.1~km.
Though use of a specific parametrization makes our results
model-dependent, we have checked that they are consistent with another common EOS
parametrization, the piecewise polytropic one~\cite{Read:2008iy,Raaijmakers:2018bln}, as also found in~\cite{PhysRevD.98.063004}.

In this analysis, we follow the methodology detailed in~\cite{PhysRevD.98.063004}, developed from the work of~\cite{Lackey:2014fwa}, to sample directly in an EOS parameter space. 
We sample uniformly in all EOS parameters within the following ranges: $\gamma_0 \in [0.2,2]$,
$\gamma_1 \in [-1.6,1.7]$, $\gamma_2 \in [-0.6,0.6]$, and $\gamma_3 \in [-0.02,0.02]$
and additionally impose that the adiabatic index $\Gamma(p) \in [0.6,4.5]$. This choice of prior ranges for the EOS
parameters was chosen such that our parametrization encompasses a wide range of candidate EOSs~\cite{PhysRevD.82.103011} and leads to NSs with a compactness below 0.33 and a tidal deformability above about 10.
Then for each sample, the four EOS parameters and the masses are mapped to a $(\Lambda_{1},\Lambda_{2})$ pair through
the Tolman-Oppenheimer-Volkoff (TOV) equations describing the equilibrium configuration of a spherical star~\cite{1983bhwd.book.....S}. The two tidal deformabilities are then used to compute the waveform template.

Sampling directly in the EOS parameter space allows for certain prior constraints
to be conveniently incorporated in the analysis.
In our analysis, we impose the following criteria on all EOS and mass samples: (i) causality, the speed 
of sound in the NS ($\sqrt{d p/d \epsilon}$) must be less than the speed of light (plus 10\% to allow for imperfect parametrization) up to the central pressure of the heaviest star supported by the EOS; (ii) internal consistency, the 
EOS must support the proposed masses of each component; and (iii) observational consistency,
the EOS must have a maximum mass at least as high as previously observed NS masses, specifically $\Maxmass \, \Msun$. Another condition the EOS must obey is that of
thermodynamic stability; the EOS must be monotonically increasing ($d\epsilon/dp > 0$). This condition is built into
the parametrization~\cite{PhysRevD.82.103011}, so we do not need to explicitly impose it.

\section{Results}

\begin{figure}[htb]
\centering
\includegraphics[width=\columnwidth]{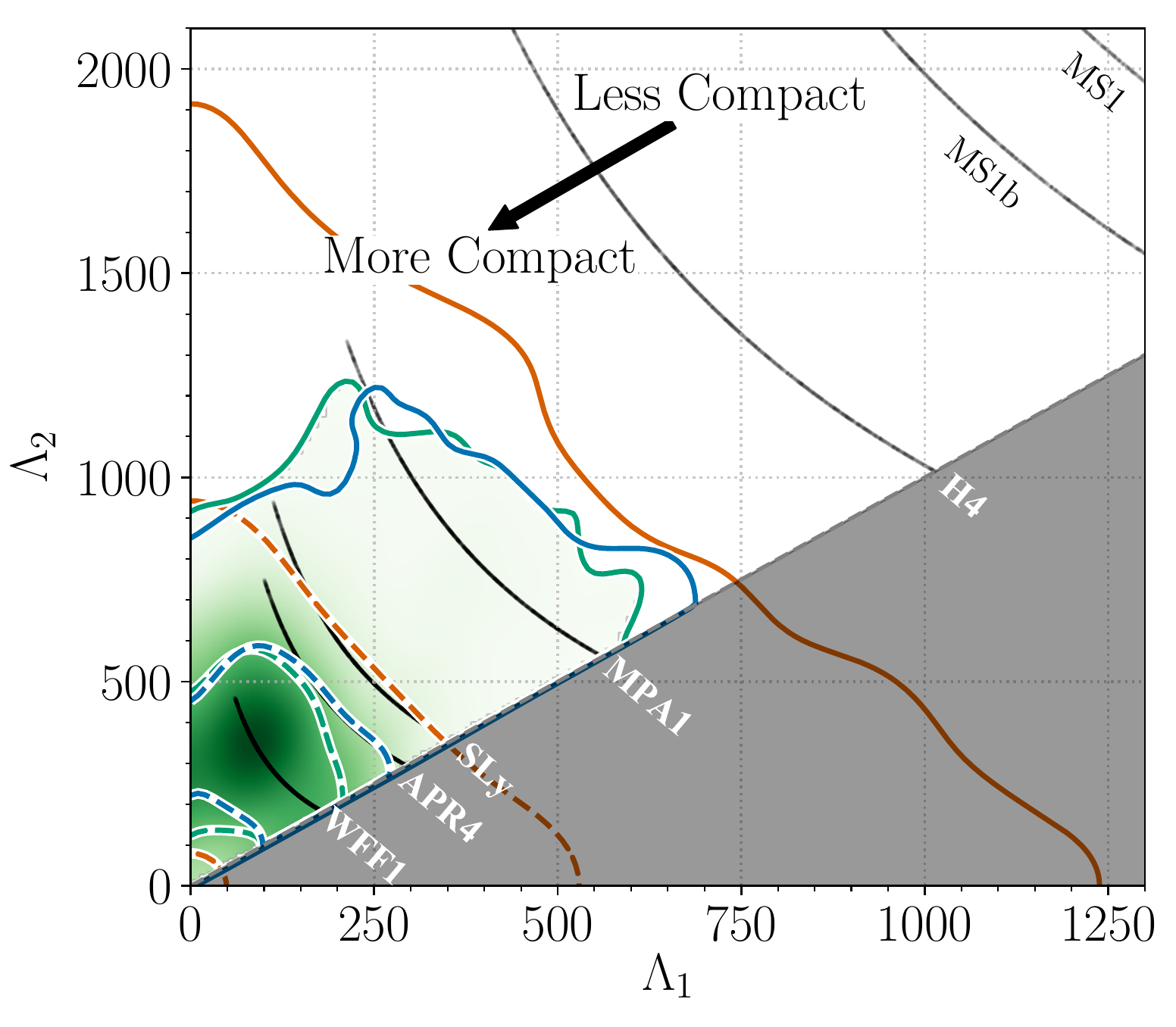}\\
\caption{Marginalized posterior for the tidal deformabilities of the two binary components of GW170817. 
The green shading shows the posterior obtained using the $\Lambda_a(\Lambda_s,q)$ EOS-insensitive relation
to impose a common EOS for the two bodies,
while the green, blue, and orange lines denote 50\% (dashed) and 90\% (solid) credible levels for
the posteriors obtained using EOS-insensitive relations, a parametrized EOS without a maximum mass requirement, and independent EOSs (taken from~\cite{Abbott:2018wiz}), respectively. The gray shading
corresponds to the unphysical region $\Lambda_2<\Lambda_1$ while the seven black scatter regions
give the tidal parameters predicted by characteristic EOS models for this
event~\cite{PhysRevC.38.1010,1998PhRvC..58.1804A,Douchin:2001sv,1987PhLB..199..469M,2006PhRvD..73b4021L,Mueller:1996pm,Read:2008iy}.}
\label{fig:l1l2_BL}
\end{figure}

We begin by demonstrating the improvement in the measurement of the tidal deformability
parameters due to imposing a common but unknown EOS for the two NSs.
In Fig.~\ref{fig:l1l2_BL} we show the marginalized joint posterior PDF for the individual tidal deformabilities.
We show results from our analysis using the $\Lambda_a(\Lambda_s,q)$ relation in green and 
the parametrized EOS without a maximum mass constraint in blue. These
are compared to results from~\cite{Abbott:2018wiz}, where the two tidal deformability parameters
are sampled independently, in orange.
The shaded region marks the $\Lambda_2<\Lambda_1$ region that is naturally
excluded when a common realistic EOS is assumed, but is not excluded from the analysis of~\cite{Abbott:2018wiz}. 
In both cases imposing a common EOS leads to a smaller uncertainty in the tidal
deformability measurement. The area of the 90\% credible region for the $\Lambda_1$--$\Lambda_2$ posterior shrinks by a factor of
$\sim \RATIOCAfromUR$, which is consistent with the results of~\cite{Chatziioannou:2018vzf} for
soft EOSs and NSs with similar masses. The tidal deformability of a
$1.4 \,\Msun$ NS can be estimated through a linear expansion of $\Lambda(m) m^5$ around $1.4\,\Msun$ as in~\cite{PhysRevLett.111.071101,PhysRevD.92.023012,DetectionPaper} to be $\Lambda_{1.4}=\Lambdameasurement$ at the $90\%$ level when a common EOS is imposed (here and throughout this paper we quote symmetric credible intervals).
Our results suggest that ``soft'' EOSs such as APR4, which predict smaller values of the
tidal deformability parameter, are favored over ``stiff'' EOSs such as H4 or MS1, which
predict larger values of the tidal deformability parameter and lie outside the $90\%$ credible region.

\begin{figure}[!htb]
\begin{centering}
\includegraphics[width=\columnwidth]{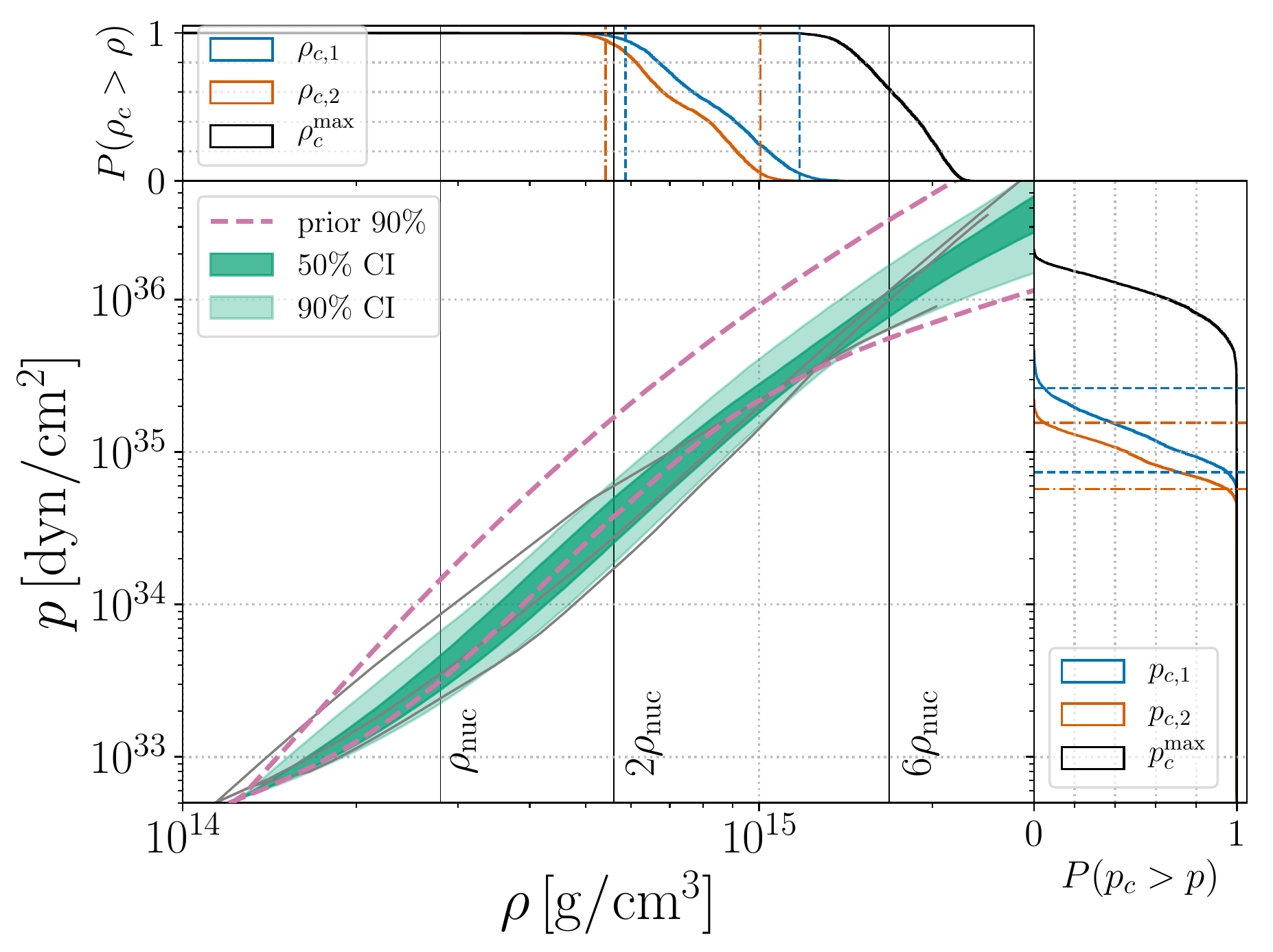}
\caption{Marginalized posterior (green bands) and prior (purple dashed) for the pressure $p$ as a function of the rest-mass
density $\rho$ of the NS interior using the spectral EOS parametrization and imposing a lower
limit on the maximum NS mass supported by the EOS of $\Maxmass \, \Msun$. The dark (light) shaded region corresponds
to the 50\% (90\%) posterior credible level and the purple dashed lines show the 90\% prior credible
interval. 
Vertical lines correspond to once, twice, and six times the nuclear saturation density.
Overplotted in gray are representative EOS models~\cite{PhysRevC.38.1010,1998PhRvC..58.1804A,2006PhRvD..73b4021L}, using data taken from~\cite{Ozel:2016oaf}; from top to bottom at $2 \rho_{\mathrm{nuc}}$ we show H4, APR4, and WFF1.
The corner plots show cumulative posteriors of central densities $\rho_c$ (top) and central pressures $p_c$ (right) for the two NSs (blue and orange),
as well as for the heaviest NS that the EOS supports (black).
The 90\% credible intervals for $\rho_c$ and $p_c$ are denoted by vertical and horizontal lines respectively 
for the heavier (blue dashed) and lighter (orange dot-dashed) NS.}
\label{fig:Prho_PW}
\end{centering}
\end{figure}

We next explore what inferences we can make about the structure of NSs.
We do this using the spectral EOS parametrization described above
in combination with the requirement that the EOS must support NSs up to at
least $\Maxmass \, \Msun$, a conservative estimate based on the heaviest known pulsar~\cite{Antoniadis:2013pzd}.
From this we obtain a posterior for the NS interior pressure as a function of rest-mass density.
The result is shown in Fig.~\ref{fig:Prho_PW}, along with marginalized posteriors for central densities and central pressures
and predictions of the pressure-density relationship from various EOS models.
The pressure posterior is shifted from the 90\% credible prior region (marked by the purple dashed lines) and towards the soft floor of the parametrized family of EOS.
This means that the posterior is indicating more support for softer EOS than the prior.
The solid vertical lines denote the nuclear saturation density
and two more rest-mass density values that are known to approximately correlate with bulk macroscopic
properties of NSs~\cite{Ozel:2016oaf}. The pressure at twice (six times) the nuclear saturation
density is measured to be $\Pattwicesaturation \times 10^{34}$ ($\Patsixtimessaturation \times 10^{35}$) $\mathrm{dyn}/\mathrm{cm}^2$ at the 90\% level.

The pressure posterior appears to show minor signs of a bend above a density of $\sim 5\rho_{\mathrm{nuc}}$. 
Evidence of such behavior at high densities would be an indication of extra degrees of freedom,
though this is not an outcome of the GW data alone.
Indeed in the top (right) panel, the vertical (horizontal) lines denote the 90\% confidence intervals for the 
central densities (pressures) of the two stars, suggesting that our data are not informative for densities (pressures) above those intervals.
The bend is an outcome of two competing effects: the GW data point toward a lower pressure, 
while the requirement that the EOS supports
masses above $\Maxmass \, \Msun$ demands a high pressure at large densities.
The result is a precise pressure estimate at around $5\rho_{\mathrm{nuc}}$ 
and a broadening above that, giving
the impression of a bend in the pressure.
We have verified that the bend is absent if we remove the maximum
mass constraint from our analysis. 

\begin{figure*}[htb]
\centering
\includegraphics[width=\columnwidth]{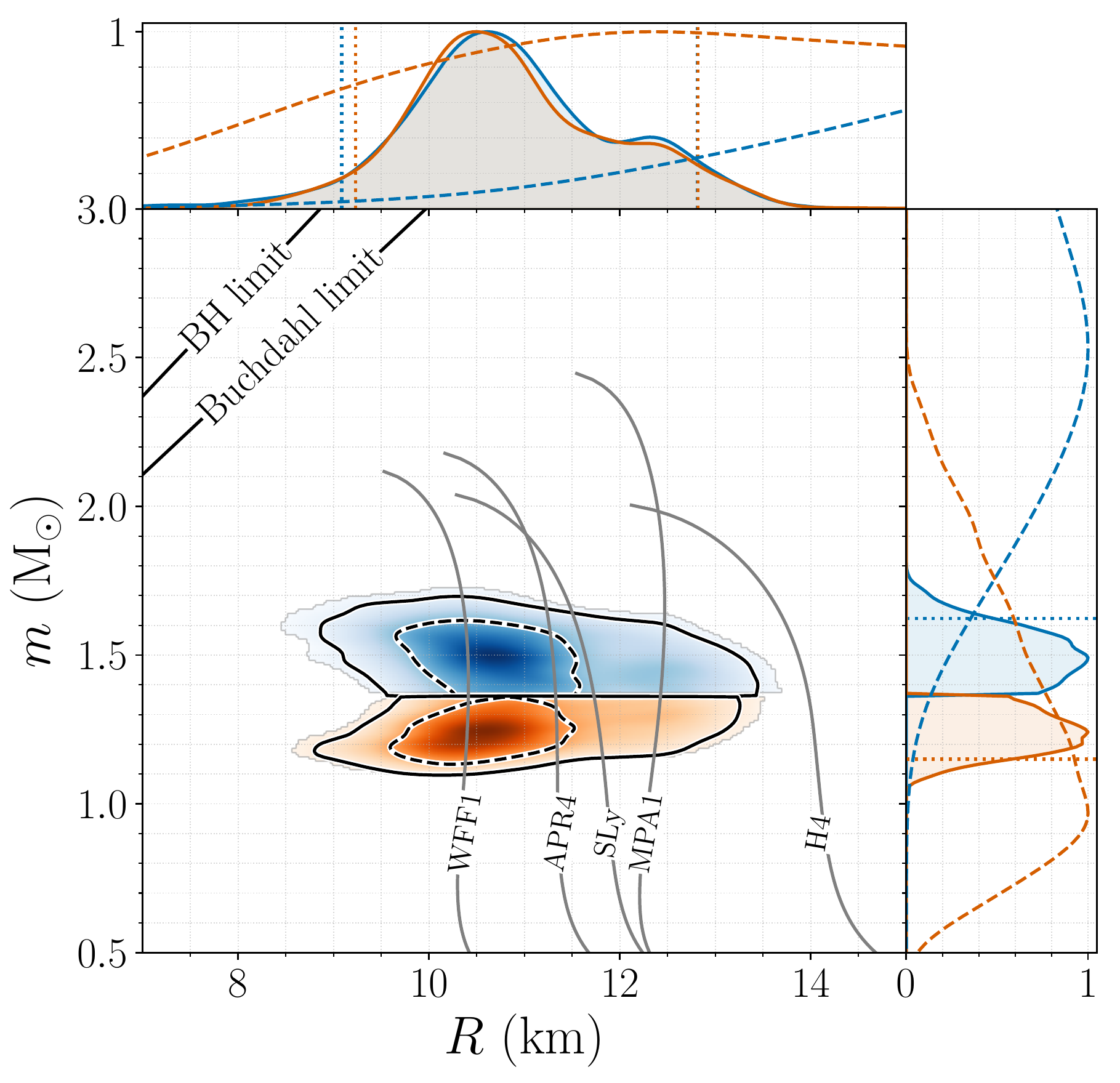}
\includegraphics[width=\columnwidth]{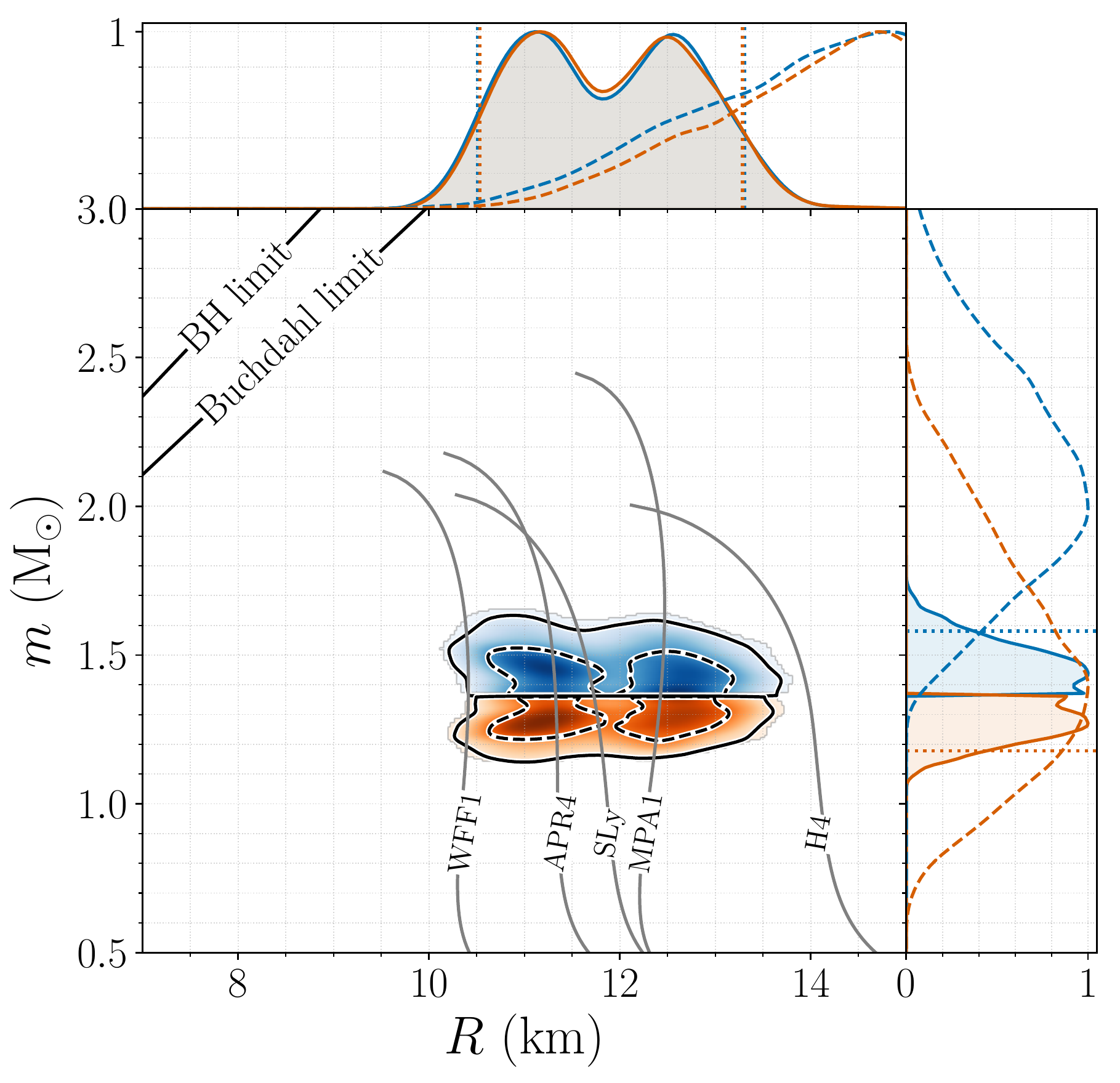}
\caption{Marginalized posterior for the mass $m$ and areal radius $R$ of each binary
component using EOS-insensitive relations (left panel) and a parametrized EOS where we
impose a lower limit on the maximum mass of $\Maxmass \, \Msun$ (right panel).
The top blue (bottom orange) posterior corresponds to the heavier (lighter) NS.
Example mass-radius curves for selected EOSs are overplotted in gray. The lines in the top left denote the
Schwarzschild BH ($R=2m$) and Buchdahl ($R=9m/4$) limits. 
In the one-dimensional plots, solid lines are used for the posteriors, while dashed
lines are used for the corresponding parameter priors. Dotted vertical lines are
used for the bounds of the 90\% credible intervals.}
\label{fig:mr}
\end{figure*}

Finally we place constraints in the 2-dimensional parameter space
of the NS mass and areal radius for each binary component.
This posterior is shown in Fig.~\ref{fig:mr}.
The left panel is obtained by first using the $\Lambda_a(\Lambda_s,q)$
relation to obtain tidal deformability samples assuming a common EOS 
and then using the $\Lambda$--$C$ relation to compute the NS radii. 
The right panel
is computed by integrating the TOV equation to compute the radius for each sample in the spectral
EOS parametrization after imposing a maximum mass of at least $\Maxmass \, \Msun$. At the 90\% level, the
radii of the two NSs are $R_1= \RadiusheavyUR$\,km and  $R_2= \RadiuslightUR$\,km from the 
left panel and $R_1= \RadiusheavyPPmm$\,km and  $R_2= \RadiuslightPPmm$\,km from the right panel. 
The one-sided 90\% lower [upper] limit on $m_2[m_1]$ is
$\massoneUR M_{\odot } [\masstwoUR M_{\odot }]$
from the left panel 
and $\massoneS M_{\odot } [\masstwoS M_{\odot }]$
from the right panel, consistent with the results of Ref.~\cite{Abbott:2018wiz}.
We note that the $\Lambda$--$C$ relation has not been established to values of $\Lambda$ less than 
$20$~\cite{Yagi:2016bkt}. In order to check the validity of our EOS-insensitive results in this regime, 
we first verify that the parametrized-EOS results without a maximum mass constraint satisfy the 
$\Lambda$--$C$ relation to the required accuracy, even for $\Lambda_1<20$. Furthermore, we find that 
our radius and mass estimates are unaffected if we discard all $\Lambda_1<10$ samples.

The difference between the two radius estimates is mainly due to different physical information
included in each analysis. The EOS-insensitive-relations analysis (left panel) is based on GW
data alone, while the parametrized-EOS analysis (right panel) imposes an additional
observational constraint, namely that the EOS must support NSs of at least $\Maxmass \, \Msun$.
This has a large effect on the radii priors as shown in the 1-dimensional plots of Fig.~\ref{fig:mr},
since small radii are typically predicted by soft EOSs, which cannot support large NS masses.
In the case of EOS-insensitive relations (left panel), the prior allows for smaller values of
the radius than in the parametrized-EOS case (right panel), something that is reflected
in the posteriors since the GW data alone cannot rule out radii below $\sim \Radiuscutoof $\,km.
Therefore the lower radius limit in the EOS-insensitive-relations analysis is determined by the
GW measurement, while in the case of the parametrized-EOS analysis it is determined by
the mass of the heaviest observed pulsar and its implications for NS radii~\cite{Antoniadis:2013pzd}.
Additionally, we verified that the parametrized-EOS analysis without the maximum mass
constraint leads to similar results to the EOS-insensitive-relations analysis.

To quantify the improvement from assuming that
both NSs obey the same EOS, we apply the $\Lambda$--$C$ relation to tidal
deformability samples calculated without assuming the $\Lambda_a(\Lambda_s,q)$ relation
(the orange posterior of Fig.~\ref{fig:l1l2_BL}) and obtain $R_1= \RadiusheavyOR$\,km and
$R_2= \RadiuslightOR$\,km at the 90\% level. This suggests that imposing a common EOS for
the two binary components leads to a reduction of the 90\% credible interval width for the radius
measurement of almost a factor of 2, from $\DeltaRlightOR$ to $\DeltaRlightUR$\,km.

\section{Discussion}
\label{sec:conclusions}

In this Letter, we complement our analysis of the tidal effects of GW170817 in~\cite{Abbott:2018wiz} with a 
targeted analysis that assumes astrophysically plausible NS spins and tidal parameters, as well as the same EOS for both NSs. 
This additional prior information enables us to measure NS radii with
an uncertainty less than $\DeltaRlightPPmm$\,km if consistency with observed pulsar masses is enforced, and
$\DeltaRlightUR$\,km using GW data alone at the 90\% credible level. 
We observe that, in both cases, the data are informative and drive the upper bounds on the NS radii and the stiffness of the EOS.
Simultaneously, the pressure at twice the nuclear saturation density is measured to be $p (2\rho_\mathrm{nuc}) = \Pattwicesaturation \times 10^{34} \,\mathrm{dyn}/\mathrm{cm}^{2}$.
Our results are consistent with x-ray binary observations
(see, e.g,~\cite{Steiner:2010,Ozel:2016oaf,Steiner:2017vmg,Nattila:2017wtj}) and suggest
that NS radii are not large. Additionally, our results can be compared to tidal 
inference based on the electromagnetic emission of GW170817~\cite{Shibata:2017xdx,Radice:2017lry,Coughlin:2018miv}.

Our results are comparable and consistent with studies that use the
tidal measurement from~\cite{DetectionPaper} to obtain bounds on NS radii.
Using our bound of $\Lambda_{1.4}<800$ (the only tidal parameter in~\cite{DetectionPaper},
which assumed a common EOS for both NSs) and different EOS parametrizations, several studies
found $R_{1.4}\lesssim13.5$\,km~\cite{PhysRevLett.120.172703,Zhou:2017pha,PhysRevLett.120.172702,Most:2018hfd}.
Reference~\cite{Raithel:2018ncd} arrives at a similar conclusion using our
$\tilde{\Lambda}<800$ constraint~\cite{DetectionPaper}
(though see~\cite{Abbott:2018wiz} for an amended $\tilde{\Lambda}$ bound) 
and the observation that
$\tilde{\Lambda}$ is almost insensitive to the binary mass ratio~\cite{Wade:2014vqa}.
Our improved estimate of $\Lambda_{1.4}=\Lambdameasurement$, and $R_1= \RadiusheavyUR$\,km and
$R_2= \RadiuslightUR$\,km for the EOS-insensitive-relation analysis is roughly consistent with
these estimates (see for example Fig. 1 of~\cite{PhysRevLett.120.172703}
and~\cite{PhysRevLett.120.172702}). If we additionally enforce the heaviest observed pulsar to be
supported by placing direct constraints on the EOS parameter space, we get further improvement in the radius measurement,
with $R_1= \RadiusheavyPPmm$\,km and $R_2= \RadiuslightPPmm$\,km.


A recent analysis of the GW170817 data was performed in De \textit{et al.}~\cite{De:2018uhw}
using the \tf{} model, imposing that the two NSs have the same radii which,
under the additional assumption that $\Lambda \propto C^{-6}$ (an alternative to the $\Lambda$--$C$ relation used here~\cite{Yagi:2016bkt}),
directly relates the two tidal deformabilities as $\Lambda_1=q^6 \Lambda_2$. 
After our paper appeared as a preprint, De \textit{et al.} obtained a revised estimate of the common NS radius $8.9\,\mathrm{km} < \hat{R} < 13.2\,\mathrm{km}$. Despite using a lower low-frequency cutoff -- and hence more data -- than our study, the result of De \textit{et al.} corresponds to a width of $4.3$\,km, which is wider than the uncertainty on radii computed under our EOS-insensitive analysis. There are differences in several details of the setup of the two analyses (most notably, frequency range, data calibration, the noise PSD estimation, waveform model, parameter priors, assumed relations between radii and $\Lambda$s and treatment of corresponding uncertainties), each of which may be responsible for part of the observed discrepancies.


Our results, and specifically the lower radius limit, do not constitute
observational proof of tidal effects in GW170817, as our analysis has explicitly assumed that the
coalescing bodies were NSs both in terms of their spins and tidal deformabilities. 
In particular, the spins are restricted to small values typical for galactic NSs in binaries, and the 
tidal deformabilites are calculated consistently assuming a common typical NS EOS.
Moreover, the $\Lambda$--$C$ map diverges as $\Lambda$ approaches zero (BH), and therefore the lower 
bounds obtained for the radii do not imply lower bounds on the tidal deformabilities.
Meanwhile, the analysis of~\cite{Abbott:2018wiz} assumes independent tidal parameters 
and finds a lower bound on $\tilde{\Lambda}$ only under the small-spin
assumption but not if spins larger than $0.05$ are allowed.

The detection of \event{} has opened new avenues in astrophysics and in the study of matter at conditions currently unattainable in terrestrial laboratories.
As the network of GW observatories expands and improves in sensitivity, we expect many more observations of BNS mergers~\cite{Aasi:2013wya}. 
Each new observation will yield additional information about the properties of NSs, and the increasing precision of our measurements will simultaneously raise new challenges.
As statistical uncertainties shrink, systematic uncertainties that are naturally introduced by our models and the underlying assumptions of our methods may begin to dominate. 
Improved waveform models and data analysis techniques are an area of active research for the GW community, and will be required to achieve our most complete understanding of these extreme systems.

Data associated with the figures in this article, including posterior samples generated using the \imrpp{} model, can be found at \href{https://dcc.ligo.org/LIGO-P1800115/public}{dcc.ligo.org/LIGO-P1800115/public}.
The GW strain data for this event are available at the Gravitational Wave Open Science Center \cite{GWOSC:GW170817}. 
This article has been assigned the document number LIGO-P1800115.

\section{Acknowledgments}
The authors gratefully acknowledge the support of the U.S.
National Science Foundation (NSF) for the construction and operation of the
LIGO Laboratory and Advanced LIGO as well as the Science and Technology Facilities Council (STFC) of the
United Kingdom, the Max-Planck-Society (MPS), and the State of
Niedersachsen/Germany for support of the construction of Advanced LIGO 
and construction and operation of the GEO600 detector. 
Additional support for Advanced LIGO was provided by the Australian Research Council.
The authors gratefully acknowledge the Italian Istituto Nazionale di Fisica Nucleare (INFN),  
the French Centre National de la Recherche Scientifique (CNRS) and
the Foundation for Fundamental Research on Matter supported by the Netherlands Organisation for Scientific Research, 
for the construction and operation of the Virgo detector
and the creation and support  of the EGO consortium. 
The authors also gratefully acknowledge research support from these agencies as well as by 
the Council of Scientific and Industrial Research of India, 
the Department of Science and Technology, India,
the Science \& Engineering Research Board (SERB), India,
the Ministry of Human Resource Development, India,
the Spanish  Agencia Estatal de Investigaci\'on,
the Vicepresid\`encia i Conselleria d'Innovaci\'o, Recerca i Turisme and the Conselleria d'Educaci\'o i Universitat del Govern de les Illes Balears,
the Conselleria d'Educaci\'o, Investigaci\'o, Cultura i Esport de la Generalitat Valenciana,
the National Science Centre of Poland,
the Swiss National Science Foundation (SNSF),
the Russian Foundation for Basic Research, 
the Russian Science Foundation,
the European Commission,
the European Regional Development Funds (ERDF),
the Royal Society, 
the Scottish Funding Council, 
the Scottish Universities Physics Alliance, 
the Hungarian Scientific Research Fund (OTKA),
the Lyon Institute of Origins (LIO),
the Paris \^{I}le-de-France Region, 
the National Research, Development and Innovation Office Hungary (NKFI), 
the National Research Foundation of Korea,
Industry Canada and the Province of Ontario through the Ministry of Economic Development and Innovation, 
the Natural Science and Engineering Research Council Canada,
the Canadian Institute for Advanced Research,
the Brazilian Ministry of Science, Technology, Innovations, and Communications,
the International Center for Theoretical Physics South American Institute for Fundamental Research (ICTP-SAIFR), 
the Research Grants Council of Hong Kong,
the National Natural Science Foundation of China (NSFC),
the Leverhulme Trust, 
the Research Corporation, 
the Ministry of Science and Technology (MOST), Taiwan
and
the Kavli Foundation.
The authors gratefully acknowledge the support of the NSF, STFC, MPS, INFN, CNRS and the
State of Niedersachsen/Germany for provision of computational resources.

The authors would like to thank N.~K.~Johnson-McDaniel, W.~Kastaun, J.~L.~Friedman, G.~Baym, J.~M.~Lattimer, L.~Rezzolla, M.~B.~Tsang and M.C.~Miller for their useful comments.

\def\url{} 
\bibliography{cbc-group.bib}

\begin{thebibliography}{131}%
\makeatletter
\providecommand \@ifxundefined [1]{%
 \@ifx{#1\undefined}
}%
\providecommand \@ifnum [1]{%
 \ifnum #1\expandafter \@firstoftwo
 \else \expandafter \@secondoftwo
 \fi
}%
\providecommand \@ifx [1]{%
 \ifx #1\expandafter \@firstoftwo
 \else \expandafter \@secondoftwo
 \fi
}%
\providecommand \natexlab [1]{#1}%
\providecommand \enquote  [1]{``#1''}%
\providecommand \bibnamefont  [1]{#1}%
\providecommand \bibfnamefont [1]{#1}%
\providecommand \citenamefont [1]{#1}%
\providecommand \href@noop [0]{\@secondoftwo}%
\providecommand \href [0]{\begingroup \@sanitize@url \@href}%
\providecommand \@href[1]{\@@startlink{#1}\@@href}%
\providecommand \@@href[1]{\endgroup#1\@@endlink}%
\providecommand \@sanitize@url [0]{\catcode `\\12\catcode `\$12\catcode
  `\&12\catcode `\#12\catcode `\^12\catcode `\_12\catcode `\%12\relax}%
\providecommand \@@startlink[1]{}%
\providecommand \@@endlink[0]{}%
\providecommand \url  [0]{\begingroup\@sanitize@url \@url }%
\providecommand \@url [1]{\endgroup\@href {#1}{\urlprefix }}%
\providecommand \urlprefix  [0]{URL }%
\providecommand \Eprint [0]{\href }%
\providecommand \doibase [0]{http://dx.doi.org/}%
\providecommand \selectlanguage [0]{\@gobble}%
\providecommand \bibinfo  [0]{\@secondoftwo}%
\providecommand \bibfield  [0]{\@secondoftwo}%
\providecommand \translation [1]{[#1]}%
\providecommand \BibitemOpen [0]{}%
\providecommand \bibitemStop [0]{}%
\providecommand \bibitemNoStop [0]{.\EOS\space}%
\providecommand \EOS [0]{\spacefactor3000\relax}%
\providecommand \BibitemShut  [1]{\csname bibitem#1\endcsname}%
\let\auto@bib@innerbib\@empty
\bibitem [{\citenamefont {Aasi}\ \emph {et~al.}(2015)\citenamefont {Aasi} \emph
  {et~al.}}]{cqg.32.074001.15}%
  \BibitemOpen
  \bibfield  {author} {\bibinfo {author} {\bibfnamefont {J.}~\bibnamefont
  {Aasi}} \emph {et~al.} (\bibinfo {collaboration} {LIGO Scientific
  Collaboration}),\ }\href {\doibase 10.1088/0264-9381/32/7/074001} {\bibfield
  {journal} {\bibinfo  {journal} {Class. Quant. Grav.}\ }\textbf {\bibinfo
  {volume} {32}},\ \bibinfo {pages} {074001} (\bibinfo {year}
  {2015})}\BibitemShut {NoStop}%
\bibitem [{\citenamefont {Acernese}\ \emph {et~al.}(2015)\citenamefont
  {Acernese} \emph {et~al.}}]{TheVirgo:2014hva}%
  \BibitemOpen
  \bibfield  {author} {\bibinfo {author} {\bibfnamefont {F.}~\bibnamefont
  {Acernese}} \emph {et~al.} (\bibinfo {collaboration} {The Virgo
  Collaboration}),\ }\href {\doibase 10.1088/0264-9381/32/2/024001} {\bibfield
  {journal} {\bibinfo  {journal} {Class. Quantum Grav.}\ }\textbf {\bibinfo
  {volume} {32}},\ \bibinfo {pages} {024001} (\bibinfo {year} {2015})},\
  \Eprint {http://arxiv.org/abs/1408.3978} {arXiv:1408.3978 [gr-qc]}
  \BibitemShut {NoStop}%
\bibitem [{\citenamefont {{Abbott}}\ \emph {et~al.}(2016)\citenamefont
  {{Abbott}}, \citenamefont {{Abbott}}, \citenamefont {{Abbott}}, \citenamefont
  {{Abernathy}}, \citenamefont {{Acernese}}, \citenamefont {{Ackley}},
  \citenamefont {{Adams}}, \citenamefont {{Adams}}, \citenamefont {{Addesso}},
  \citenamefont {{Adhikari}},\ and\ \citenamefont {et~al.}}]{GW150914_paper}%
  \BibitemOpen
  \bibfield  {author} {\bibinfo {author} {\bibfnamefont {B.~P.}\ \bibnamefont
  {{Abbott}}}, \bibinfo {author} {\bibfnamefont {R.}~\bibnamefont {{Abbott}}},
  \bibinfo {author} {\bibfnamefont {T.~D.}\ \bibnamefont {{Abbott}}}, \bibinfo
  {author} {\bibfnamefont {M.~R.}\ \bibnamefont {{Abernathy}}}, \bibinfo
  {author} {\bibfnamefont {F.}~\bibnamefont {{Acernese}}}, \bibinfo {author}
  {\bibfnamefont {K.}~\bibnamefont {{Ackley}}}, \bibinfo {author}
  {\bibfnamefont {C.}~\bibnamefont {{Adams}}}, \bibinfo {author} {\bibfnamefont
  {T.}~\bibnamefont {{Adams}}}, \bibinfo {author} {\bibfnamefont
  {P.}~\bibnamefont {{Addesso}}}, \bibinfo {author} {\bibfnamefont {R.~X.}\
  \bibnamefont {{Adhikari}}}, \ and\ \bibinfo {author} {\bibnamefont
  {et~al.}},\ }\href {\doibase 10.1103/PhysRevLett.116.061102} {\bibfield
  {journal} {\bibinfo  {journal} {Physical Review Letters}\ }\textbf {\bibinfo
  {volume} {116}},\ \bibinfo {eid} {061102} (\bibinfo {year} {2016})},\ \Eprint
  {http://arxiv.org/abs/1602.03837} {arXiv:1602.03837 [gr-qc]} \BibitemShut
  {NoStop}%
\bibitem [{\citenamefont {Abbott}\ \emph
  {et~al.}(2018{\natexlab{a}})\citenamefont {Abbott} \emph
  {et~al.}}]{Aasi:2013wya}%
  \BibitemOpen
  \bibfield  {author} {\bibinfo {author} {\bibfnamefont {B.~P.}\ \bibnamefont
  {Abbott}} \emph {et~al.} (\bibinfo {collaboration} {LIGO Scientific
  Collaboration, Virgo Collaboration, KAGRA Collaboration}),\ }\href {\doibase
  10.1007/s41114-018-0012-9} {\bibfield  {journal} {\bibinfo  {journal} {Living
  Rev. Rel.}\ }\textbf {\bibinfo {volume} {21}},\ \bibinfo {pages} {3}
  (\bibinfo {year} {2018}{\natexlab{a}})},\ \Eprint
  {http://arxiv.org/abs/1304.0670} {arXiv:1304.0670 [gr-qc]} \BibitemShut
  {NoStop}%
\bibitem [{\citenamefont {Abbott}\ \emph
  {et~al.}(2017{\natexlab{a}})\citenamefont {Abbott} \emph
  {et~al.}}]{DetectionPaper}%
  \BibitemOpen
  \bibfield  {author} {\bibinfo {author} {\bibfnamefont {B.}~\bibnamefont
  {Abbott}} \emph {et~al.} (\bibinfo {collaboration} {LIGO Scientific
  Collaboration, Virgo Collaboration}),\ }\href {\doibase
  10.1103/PhysRevLett.119.161101} {\bibfield  {journal} {\bibinfo  {journal}
  {Phys. Rev. Lett.}\ }\textbf {\bibinfo {volume} {119}},\ \bibinfo {pages}
  {161101} (\bibinfo {year} {2017}{\natexlab{a}})},\ \Eprint
  {http://arxiv.org/abs/1710.05832} {arXiv:1710.05832 [gr-qc]} \BibitemShut
  {NoStop}%
\bibitem [{\citenamefont {Abbott}\ \emph
  {et~al.}(2017{\natexlab{b}})\citenamefont {Abbott} \emph
  {et~al.}}]{Monitor:2017mdv}%
  \BibitemOpen
  \bibfield  {author} {\bibinfo {author} {\bibfnamefont {B.~P.}\ \bibnamefont
  {Abbott}} \emph {et~al.} (\bibinfo {collaboration} {Virgo, Fermi-GBM,
  INTEGRAL, LIGO Scientific Collaboration}),\ }\href {\doibase
  10.3847/2041-8213/aa920c} {\bibfield  {journal} {\bibinfo  {journal}
  {Astrophys. J.}\ }\textbf {\bibinfo {volume} {848}},\ \bibinfo {pages} {L13}
  (\bibinfo {year} {2017}{\natexlab{b}})},\ \Eprint
  {http://arxiv.org/abs/1710.05834} {arXiv:1710.05834 [astro-ph.HE]}
  \BibitemShut {NoStop}%
\bibitem [{\citenamefont {Goldstein}\ \emph {et~al.}(2017)\citenamefont
  {Goldstein} \emph {et~al.}}]{Goldstein:2017mmi}%
  \BibitemOpen
  \bibfield  {author} {\bibinfo {author} {\bibfnamefont {A.}~\bibnamefont
  {Goldstein}} \emph {et~al.},\ }\href {\doibase 10.3847/2041-8213/aa8f41}
  {\bibfield  {journal} {\bibinfo  {journal} {Astrophys. J.}\ }\textbf
  {\bibinfo {volume} {848}},\ \bibinfo {pages} {L14} (\bibinfo {year}
  {2017})},\ \Eprint {http://arxiv.org/abs/1710.05446} {arXiv:1710.05446
  [astro-ph.HE]} \BibitemShut {NoStop}%
\bibitem [{\citenamefont {{Abbott}}\ \emph
  {et~al.}(2017{\natexlab{a}})\citenamefont {{Abbott}} \emph
  {et~al.}}]{GBM:2017lvd}%
  \BibitemOpen
  \bibfield  {author} {\bibinfo {author} {\bibfnamefont {B.~P.}\ \bibnamefont
  {{Abbott}}} \emph {et~al.} (\bibinfo {collaboration} {GROND, SALT Group,
  OzGrav, CAASTROs, DFN, DES, INTEGRAL, Virgo, Insight-HXMT, MAXI Team, J-GEM,
  RATIR, ATLAS, IceCube, LWA, ePESSTO, GRAWITA, RIMAS, SKA South
  Africa/MeerKAT, H.E.S.S., Fermi Large Area Telescope, 1M2H Team, IKI-GW
  Follow-up, Fermi GBM, Pi of Sky, DWF (Deeper Wider Faster Program), MASTER,
  AstroSat Cadmium Zinc Telluride Imager Team, Swift, Pierre Auger, ASKAP,
  VINROUGE, JAGWAR, Chandra Team at McGill University, TTU-NRAO, GROWTH, AGILE
  Team, MWA, ATCA, AST3, TOROS, Pan-STARRS, NuSTAR, BOOTES, CaltechNRAO, LIGO
  Scientific Collaboration, High Time Resolution Universe Survey, Nordic
  Optical Telescope, Las Cumbres Observatory Group, TZAC Consortium, LOFAR,
  IPN, DLT40, Texas Tech University, HAWC, ANTARES, KU, Dark Energy Camera
  GW-EM, CALET, Euro VLBI Team, ALMA}),\ }\href {\doibase
  10.3847/2041-8213/aa91c9} {\bibfield  {journal} {\bibinfo  {journal}
  {Astrophys. J.}\ }\textbf {\bibinfo {volume} {848}},\ \bibinfo {pages} {L12}
  (\bibinfo {year} {2017}{\natexlab{a}})},\ \Eprint
  {http://arxiv.org/abs/1710.05833} {arXiv:1710.05833 [astro-ph.HE]}
  \BibitemShut {NoStop}%
\bibitem [{\citenamefont {Coulter}\ \emph {et~al.}(2017)\citenamefont {Coulter}
  \emph {et~al.}}]{Coulter:2017wya}%
  \BibitemOpen
  \bibfield  {author} {\bibinfo {author} {\bibfnamefont {D.~A.}\ \bibnamefont
  {Coulter}} \emph {et~al.},\ }\href {\doibase 10.1126/science.aap9811}
  {\bibfield  {journal} {\bibinfo  {journal} {Science}\ }\textbf {\bibinfo
  {volume} {358}},\ \bibinfo {pages} {1556} (\bibinfo {year} {2017})},\ \Eprint
  {http://arxiv.org/abs/1710.05452} {arXiv:1710.05452 [astro-ph.HE]}
  \BibitemShut {NoStop}%
\bibitem [{\citenamefont {Troja}\ \emph {et~al.}(2017)\citenamefont {Troja}
  \emph {et~al.}}]{Troja:2017nqp}%
  \BibitemOpen
  \bibfield  {author} {\bibinfo {author} {\bibfnamefont {E.}~\bibnamefont
  {Troja}} \emph {et~al.},\ }\href {\doibase 10.1038/nature24290} {\bibfield
  {journal} {\bibinfo  {journal} {Nature}\ }\textbf {\bibinfo {volume} {551}},\
  \bibinfo {pages} {71} (\bibinfo {year} {2017})},\ \Eprint
  {http://arxiv.org/abs/1710.05433} {arXiv:1710.05433 [astro-ph.HE]}
  \BibitemShut {NoStop}%
\bibitem [{\citenamefont {Haggard}\ \emph {et~al.}(2017)\citenamefont
  {Haggard}, \citenamefont {Nynka}, \citenamefont {Ruan}, \citenamefont
  {Kalogera}, \citenamefont {Bradley~Cenko}, \citenamefont {Evans},\ and\
  \citenamefont {Kennea}}]{Haggard:2017qne}%
  \BibitemOpen
  \bibfield  {author} {\bibinfo {author} {\bibfnamefont {D.}~\bibnamefont
  {Haggard}}, \bibinfo {author} {\bibfnamefont {M.}~\bibnamefont {Nynka}},
  \bibinfo {author} {\bibfnamefont {J.~J.}\ \bibnamefont {Ruan}}, \bibinfo
  {author} {\bibfnamefont {V.}~\bibnamefont {Kalogera}}, \bibinfo {author}
  {\bibfnamefont {S.}~\bibnamefont {Bradley~Cenko}}, \bibinfo {author}
  {\bibfnamefont {P.}~\bibnamefont {Evans}}, \ and\ \bibinfo {author}
  {\bibfnamefont {J.~A.}\ \bibnamefont {Kennea}},\ }\href {\doibase
  10.3847/2041-8213/aa8ede} {\bibfield  {journal} {\bibinfo  {journal}
  {Astrophys. J.}\ }\textbf {\bibinfo {volume} {848}},\ \bibinfo {pages} {L25}
  (\bibinfo {year} {2017})},\ \Eprint {http://arxiv.org/abs/1710.05852}
  {arXiv:1710.05852 [astro-ph.HE]} \BibitemShut {NoStop}%
\bibitem [{\citenamefont {Hallinan}\ \emph {et~al.}(2017)\citenamefont
  {Hallinan} \emph {et~al.}}]{Hallinan:2017woc}%
  \BibitemOpen
  \bibfield  {author} {\bibinfo {author} {\bibfnamefont {G.}~\bibnamefont
  {Hallinan}} \emph {et~al.},\ }\href {\doibase 10.1126/science.aap9855}
  {\bibfield  {journal} {\bibinfo  {journal} {Science}\ }\textbf {\bibinfo
  {volume} {358}},\ \bibinfo {pages} {1579} (\bibinfo {year} {2017})},\ \Eprint
  {http://arxiv.org/abs/1710.05435} {arXiv:1710.05435 [astro-ph.HE]}
  \BibitemShut {NoStop}%
\bibitem [{\citenamefont {{Steiner}}\ \emph {et~al.}(2015)\citenamefont
  {{Steiner}}, \citenamefont {{Gandolfi}}, \citenamefont {{Fattoyev}},\ and\
  \citenamefont {{Newton}}}]{2015PhRvC..91a5804S}%
  \BibitemOpen
  \bibfield  {author} {\bibinfo {author} {\bibfnamefont {A.~W.}\ \bibnamefont
  {{Steiner}}}, \bibinfo {author} {\bibfnamefont {S.}~\bibnamefont
  {{Gandolfi}}}, \bibinfo {author} {\bibfnamefont {F.~J.}\ \bibnamefont
  {{Fattoyev}}}, \ and\ \bibinfo {author} {\bibfnamefont {W.~G.}\ \bibnamefont
  {{Newton}}},\ }\href {\doibase 10.1103/PhysRevC.91.015804} {\bibfield
  {journal} {\bibinfo  {journal} {\prc}\ }\textbf {\bibinfo {volume} {91}},\
  \bibinfo {eid} {015804} (\bibinfo {year} {2015})},\ \Eprint
  {http://arxiv.org/abs/1403.7546} {arXiv:1403.7546 [nucl-th]} \BibitemShut
  {NoStop}%
\bibitem [{\citenamefont {{Tsang}}\ \emph {et~al.}(2012)\citenamefont
  {{Tsang}}, \citenamefont {{Stone}}, \citenamefont {{Camera}}, \citenamefont
  {{Danielewicz}}, \citenamefont {{Gandolfi}}, \citenamefont {{Hebeler}},
  \citenamefont {{Horowitz}}, \citenamefont {{Lee}}, \citenamefont {{Lynch}},
  \citenamefont {{Kohley}}, \citenamefont {{Lemmon}}, \citenamefont
  {{M{\"o}ller}}, \citenamefont {{Murakami}}, \citenamefont {{Riordan}},
  \citenamefont {{Roca-Maza}}, \citenamefont {{Sammarruca}}, \citenamefont
  {{Steiner}}, \citenamefont {{Vida{\~n}a}},\ and\ \citenamefont
  {{Yennello}}}]{2012PhRvC..86a5803T}%
  \BibitemOpen
  \bibfield  {author} {\bibinfo {author} {\bibfnamefont {M.~B.}\ \bibnamefont
  {{Tsang}}}, \bibinfo {author} {\bibfnamefont {J.~R.}\ \bibnamefont
  {{Stone}}}, \bibinfo {author} {\bibfnamefont {F.}~\bibnamefont {{Camera}}},
  \bibinfo {author} {\bibfnamefont {P.}~\bibnamefont {{Danielewicz}}}, \bibinfo
  {author} {\bibfnamefont {S.}~\bibnamefont {{Gandolfi}}}, \bibinfo {author}
  {\bibfnamefont {K.}~\bibnamefont {{Hebeler}}}, \bibinfo {author}
  {\bibfnamefont {C.~J.}\ \bibnamefont {{Horowitz}}}, \bibinfo {author}
  {\bibfnamefont {J.}~\bibnamefont {{Lee}}}, \bibinfo {author} {\bibfnamefont
  {W.~G.}\ \bibnamefont {{Lynch}}}, \bibinfo {author} {\bibfnamefont
  {Z.}~\bibnamefont {{Kohley}}}, \bibinfo {author} {\bibfnamefont
  {R.}~\bibnamefont {{Lemmon}}}, \bibinfo {author} {\bibfnamefont
  {P.}~\bibnamefont {{M{\"o}ller}}}, \bibinfo {author} {\bibfnamefont
  {T.}~\bibnamefont {{Murakami}}}, \bibinfo {author} {\bibfnamefont
  {S.}~\bibnamefont {{Riordan}}}, \bibinfo {author} {\bibfnamefont
  {X.}~\bibnamefont {{Roca-Maza}}}, \bibinfo {author} {\bibfnamefont
  {F.}~\bibnamefont {{Sammarruca}}}, \bibinfo {author} {\bibfnamefont {A.~W.}\
  \bibnamefont {{Steiner}}}, \bibinfo {author} {\bibfnamefont {I.}~\bibnamefont
  {{Vida{\~n}a}}}, \ and\ \bibinfo {author} {\bibfnamefont {S.~J.}\
  \bibnamefont {{Yennello}}},\ }\href {\doibase 10.1103/PhysRevC.86.015803}
  {\bibfield  {journal} {\bibinfo  {journal} {\prc}\ }\textbf {\bibinfo
  {volume} {86}},\ \bibinfo {eid} {015803} (\bibinfo {year} {2012})},\ \Eprint
  {http://arxiv.org/abs/1204.0466} {arXiv:1204.0466 [nucl-ex]} \BibitemShut
  {NoStop}%
\bibitem [{\citenamefont {{Baldo}}\ and\ \citenamefont
  {{Burgio}}(2016)}]{2016PrPNP..91..203B}%
  \BibitemOpen
  \bibfield  {author} {\bibinfo {author} {\bibfnamefont {M.}~\bibnamefont
  {{Baldo}}}\ and\ \bibinfo {author} {\bibfnamefont {G.~F.}\ \bibnamefont
  {{Burgio}}},\ }\href {\doibase 10.1016/j.ppnp.2016.06.006} {\bibfield
  {journal} {\bibinfo  {journal} {Progress in Particle and Nuclear Physics}\
  }\textbf {\bibinfo {volume} {91}},\ \bibinfo {pages} {203} (\bibinfo {year}
  {2016})},\ \Eprint {http://arxiv.org/abs/1606.08838} {arXiv:1606.08838
  [nucl-th]} \BibitemShut {NoStop}%
\bibitem [{\citenamefont {{Lattimer}}\ and\ \citenamefont
  {{Prakash}}(2016)}]{2016PhR...621..127L}%
  \BibitemOpen
  \bibfield  {author} {\bibinfo {author} {\bibfnamefont {J.~M.}\ \bibnamefont
  {{Lattimer}}}\ and\ \bibinfo {author} {\bibfnamefont {M.}~\bibnamefont
  {{Prakash}}},\ }\href {\doibase 10.1016/j.physrep.2015.12.005} {\bibfield
  {journal} {\bibinfo  {journal} {Phys. Rep.}\ }\textbf {\bibinfo {volume}
  {621}},\ \bibinfo {pages} {127} (\bibinfo {year} {2016})},\ \Eprint
  {http://arxiv.org/abs/1512.07820} {arXiv:1512.07820 [astro-ph.SR]}
  \BibitemShut {NoStop}%
\bibitem [{\citenamefont {{Oertel}}\ \emph {et~al.}(2017)\citenamefont
  {{Oertel}}, \citenamefont {{Hempel}}, \citenamefont {{Kl{\"a}hn}},\ and\
  \citenamefont {{Typel}}}]{2017RvMP...89a5007O}%
  \BibitemOpen
  \bibfield  {author} {\bibinfo {author} {\bibfnamefont {M.}~\bibnamefont
  {{Oertel}}}, \bibinfo {author} {\bibfnamefont {M.}~\bibnamefont {{Hempel}}},
  \bibinfo {author} {\bibfnamefont {T.}~\bibnamefont {{Kl{\"a}hn}}}, \ and\
  \bibinfo {author} {\bibfnamefont {S.}~\bibnamefont {{Typel}}},\ }\href
  {\doibase 10.1103/RevModPhys.89.015007} {\bibfield  {journal} {\bibinfo
  {journal} {Reviews of Modern Physics}\ }\textbf {\bibinfo {volume} {89}},\
  \bibinfo {eid} {015007} (\bibinfo {year} {2017})},\ \Eprint
  {http://arxiv.org/abs/1610.03361} {arXiv:1610.03361 [astro-ph.HE]}
  \BibitemShut {NoStop}%
\bibitem [{\citenamefont {{Hebeler}}\ \emph {et~al.}(2013)\citenamefont
  {{Hebeler}}, \citenamefont {{Lattimer}}, \citenamefont {{Pethick}},\ and\
  \citenamefont {{Schwenk}}}]{2013ApJ...773...11H}%
  \BibitemOpen
  \bibfield  {author} {\bibinfo {author} {\bibfnamefont {K.}~\bibnamefont
  {{Hebeler}}}, \bibinfo {author} {\bibfnamefont {J.~M.}\ \bibnamefont
  {{Lattimer}}}, \bibinfo {author} {\bibfnamefont {C.~J.}\ \bibnamefont
  {{Pethick}}}, \ and\ \bibinfo {author} {\bibfnamefont {A.}~\bibnamefont
  {{Schwenk}}},\ }\href {\doibase 10.1088/0004-637X/773/1/11} {\bibfield
  {journal} {\bibinfo  {journal} {\apj}\ }\textbf {\bibinfo {volume} {773}},\
  \bibinfo {eid} {11} (\bibinfo {year} {2013})},\ \Eprint
  {http://arxiv.org/abs/1303.4662} {arXiv:1303.4662 [astro-ph.SR]} \BibitemShut
  {NoStop}%
\bibitem [{\citenamefont {{\"{O}}zel}\ and\ \citenamefont
  {Freire}(2016)}]{Ozel:2016oaf}%
  \BibitemOpen
  \bibfield  {author} {\bibinfo {author} {\bibfnamefont {F.}~\bibnamefont
  {{\"{O}}zel}}\ and\ \bibinfo {author} {\bibfnamefont {P.}~\bibnamefont
  {Freire}},\ }\href {\doibase 10.1146/annurev-astro-081915-023322} {\bibfield
  {journal} {\bibinfo  {journal} {Ann. Rev. Astron. Astrophys.}\ }\textbf
  {\bibinfo {volume} {54}},\ \bibinfo {pages} {401} (\bibinfo {year} {2016})},\
  \Eprint {http://arxiv.org/abs/1603.02698} {arXiv:1603.02698 [astro-ph.HE]}
  \BibitemShut {NoStop}%
\bibitem [{\citenamefont {Steiner}\ \emph {et~al.}(2018)\citenamefont
  {Steiner}, \citenamefont {Heinke}, \citenamefont {Bogdanov}, \citenamefont
  {Li}, \citenamefont {Ho}, \citenamefont {Bahramian},\ and\ \citenamefont
  {Han}}]{Steiner:2017vmg}%
  \BibitemOpen
  \bibfield  {author} {\bibinfo {author} {\bibfnamefont {A.~W.}\ \bibnamefont
  {Steiner}}, \bibinfo {author} {\bibfnamefont {C.~O.}\ \bibnamefont {Heinke}},
  \bibinfo {author} {\bibfnamefont {S.}~\bibnamefont {Bogdanov}}, \bibinfo
  {author} {\bibfnamefont {C.}~\bibnamefont {Li}}, \bibinfo {author}
  {\bibfnamefont {W.~C.~G.}\ \bibnamefont {Ho}}, \bibinfo {author}
  {\bibfnamefont {A.}~\bibnamefont {Bahramian}}, \ and\ \bibinfo {author}
  {\bibfnamefont {S.}~\bibnamefont {Han}},\ }\href {\doibase
  10.1093/mnras/sty215} {\bibfield  {journal} {\bibinfo  {journal} {Mon. Not.
  Roy. Astron. Soc.}\ }\textbf {\bibinfo {volume} {476}},\ \bibinfo {pages}
  {421} (\bibinfo {year} {2018})},\ \Eprint {http://arxiv.org/abs/1709.05013}
  {arXiv:1709.05013 [astro-ph.HE]} \BibitemShut {NoStop}%
\bibitem [{\citenamefont {Flanagan}\ and\ \citenamefont
  {Hinderer}(2008)}]{flanagan:021502}%
  \BibitemOpen
  \bibfield  {author} {\bibinfo {author} {\bibfnamefont {E.~E.}\ \bibnamefont
  {Flanagan}}\ and\ \bibinfo {author} {\bibfnamefont {T.}~\bibnamefont
  {Hinderer}},\ }\href {\doibase 10.1103/PhysRevD.77.021502} {\bibfield
  {journal} {\bibinfo  {journal} {\prd}\ }\textbf {\bibinfo {volume} {77}},\
  \bibinfo {eid} {021502} (\bibinfo {year} {2008})}\BibitemShut {NoStop}%
\bibitem [{\citenamefont {Vines}\ \emph {et~al.}(2011)\citenamefont {Vines},
  \citenamefont {Flanagan},\ and\ \citenamefont {Hinderer}}]{Vines:2011ud}%
  \BibitemOpen
  \bibfield  {author} {\bibinfo {author} {\bibfnamefont {J.}~\bibnamefont
  {Vines}}, \bibinfo {author} {\bibfnamefont {E.~E.}\ \bibnamefont {Flanagan}},
  \ and\ \bibinfo {author} {\bibfnamefont {T.}~\bibnamefont {Hinderer}},\
  }\href {\doibase 10.1103/PhysRevD.83.084051} {\bibfield  {journal} {\bibinfo
  {journal} {\prd}\ }\textbf {\bibinfo {volume} {83}},\ \bibinfo {pages}
  {084051} (\bibinfo {year} {2011})},\ \Eprint {http://arxiv.org/abs/1101.1673}
  {arXiv:1101.1673 [gr-qc]} \BibitemShut {NoStop}%
\bibitem [{\citenamefont {Damour}\ \emph {et~al.}(2012)\citenamefont {Damour},
  \citenamefont {Nagar},\ and\ \citenamefont {Villain}}]{PhysRevD.85.123007}%
  \BibitemOpen
  \bibfield  {author} {\bibinfo {author} {\bibfnamefont {T.}~\bibnamefont
  {Damour}}, \bibinfo {author} {\bibfnamefont {A.}~\bibnamefont {Nagar}}, \
  and\ \bibinfo {author} {\bibfnamefont {L.}~\bibnamefont {Villain}},\
  }\href@noop {} {\bibfield  {journal} {\bibinfo  {journal} {Phys. Rev. D}\
  }\textbf {\bibinfo {volume} {85}},\ \bibinfo {pages} {123007} (\bibinfo
  {year} {2012})},\ \Eprint {http://arxiv.org/abs/1203.4352} {arXiv:1203.4352
  [gr-qc]} \BibitemShut {NoStop}%
\bibitem [{\citenamefont {Buonanno}\ \emph {et~al.}(2009)\citenamefont
  {Buonanno}, \citenamefont {Iyer}, \citenamefont {Ochsner}, \citenamefont
  {Pan},\ and\ \citenamefont {Sathyaprakash}}]{Buonanno:2009zt}%
  \BibitemOpen
  \bibfield  {author} {\bibinfo {author} {\bibfnamefont {A.}~\bibnamefont
  {Buonanno}}, \bibinfo {author} {\bibfnamefont {B.~R.}\ \bibnamefont {Iyer}},
  \bibinfo {author} {\bibfnamefont {E.}~\bibnamefont {Ochsner}}, \bibinfo
  {author} {\bibfnamefont {Y.}~\bibnamefont {Pan}}, \ and\ \bibinfo {author}
  {\bibfnamefont {B.~S.}\ \bibnamefont {Sathyaprakash}},\ }\href {\doibase
  10.1103/PhysRevD.80.084043} {\bibfield  {journal} {\bibinfo  {journal}
  {\prd}\ }\textbf {\bibinfo {volume} {80}},\ \bibinfo {pages} {084043}
  (\bibinfo {year} {2009})},\ \Eprint {http://arxiv.org/abs/0907.0700}
  {arXiv:0907.0700 [gr-qc]} \BibitemShut {NoStop}%
\bibitem [{\citenamefont {Blanchet}(2014)}]{Blanchet:2013haa}%
  \BibitemOpen
  \bibfield  {author} {\bibinfo {author} {\bibfnamefont {L.}~\bibnamefont
  {Blanchet}},\ }\href {\doibase 10.12942/lrr-2014-2} {\bibfield  {journal}
  {\bibinfo  {journal} {Living Rev. Rel.}\ }\textbf {\bibinfo {volume} {17}},\
  \bibinfo {pages} {2} (\bibinfo {year} {2014})},\ \Eprint
  {http://arxiv.org/abs/1310.1528} {arXiv:1310.1528 [gr-qc]} \BibitemShut
  {NoStop}%
\bibitem [{\citenamefont {Goldberger}\ and\ \citenamefont
  {Rothstein}(2006)}]{Goldberger:2004jt}%
  \BibitemOpen
  \bibfield  {author} {\bibinfo {author} {\bibfnamefont {W.~D.}\ \bibnamefont
  {Goldberger}}\ and\ \bibinfo {author} {\bibfnamefont {I.~Z.}\ \bibnamefont
  {Rothstein}},\ }\href {\doibase 10.1103/PhysRevD.73.104029} {\bibfield
  {journal} {\bibinfo  {journal} {\prd}\ }\textbf {\bibinfo {volume} {73}},\
  \bibinfo {pages} {104029} (\bibinfo {year} {2006})},\ \Eprint
  {http://arxiv.org/abs/hep-th/0409156} {arXiv:hep-th/0409156 [hep-th]}
  \BibitemShut {NoStop}%
\bibitem [{\citenamefont {Goldberger}(2007)}]{Goldberger:2007hy}%
  \BibitemOpen
  \bibfield  {author} {\bibinfo {author} {\bibfnamefont {W.~D.}\ \bibnamefont
  {Goldberger}},\ }in\ \href@noop {} {\emph {\bibinfo {booktitle} {{Les Houches
  Summer School - Session 86: Particle Physics and Cosmology: The Fabric of
  Spacetime Les Houches, France, July 31-August 25, 2006}}}}\ (\bibinfo {year}
  {2007})\ \Eprint {http://arxiv.org/abs/hep-ph/0701129} {arXiv:hep-ph/0701129
  [hep-ph]} \BibitemShut {NoStop}%
\bibitem [{\citenamefont {Damour}\ \emph
  {et~al.}(2001{\natexlab{a}})\citenamefont {Damour}, \citenamefont {Iyer},\
  and\ \citenamefont {Sathyaprakash}}]{Damour:2000zb}%
  \BibitemOpen
  \bibfield  {author} {\bibinfo {author} {\bibfnamefont {T.}~\bibnamefont
  {Damour}}, \bibinfo {author} {\bibfnamefont {B.~R.}\ \bibnamefont {Iyer}}, \
  and\ \bibinfo {author} {\bibfnamefont {B.~S.}\ \bibnamefont
  {Sathyaprakash}},\ }\href@noop {} {\bibfield  {journal} {\bibinfo  {journal}
  {\prd}\ }\textbf {\bibinfo {volume} {63}},\ \bibinfo {pages} {044023}
  (\bibinfo {year} {2001}{\natexlab{a}})}\BibitemShut {NoStop}%
\bibitem [{\citenamefont {Blanchet}\ \emph {et~al.}(1995)\citenamefont
  {Blanchet}, \citenamefont {Damour}, \citenamefont {Iyer}, \citenamefont
  {Will},\ and\ \citenamefont {Wiseman}}]{Blanchet:1995ez}%
  \BibitemOpen
  \bibfield  {author} {\bibinfo {author} {\bibfnamefont {L.}~\bibnamefont
  {Blanchet}}, \bibinfo {author} {\bibfnamefont {T.}~\bibnamefont {Damour}},
  \bibinfo {author} {\bibfnamefont {B.~R.}\ \bibnamefont {Iyer}}, \bibinfo
  {author} {\bibfnamefont {C.~M.}\ \bibnamefont {Will}}, \ and\ \bibinfo
  {author} {\bibfnamefont {A.~G.}\ \bibnamefont {Wiseman}},\ }\href@noop {}
  {\bibfield  {journal} {\bibinfo  {journal} {\prl}\ }\textbf {\bibinfo
  {volume} {74}},\ \bibinfo {pages} {3515} (\bibinfo {year} {1995})},\ \Eprint
  {http://arxiv.org/abs/9501027} {arXiv:9501027 [gr-qc]} \BibitemShut {NoStop}%
\bibitem [{\citenamefont {Blanchet}\ \emph {et~al.}(2004)\citenamefont
  {Blanchet}, \citenamefont {Damour}, \citenamefont {Esposito-Far\`ese},\ and\
  \citenamefont {Iyer}}]{Blanchet:2004ek}%
  \BibitemOpen
  \bibfield  {author} {\bibinfo {author} {\bibfnamefont {L.}~\bibnamefont
  {Blanchet}}, \bibinfo {author} {\bibfnamefont {T.}~\bibnamefont {Damour}},
  \bibinfo {author} {\bibfnamefont {G.}~\bibnamefont {Esposito-Far\`ese}}, \
  and\ \bibinfo {author} {\bibfnamefont {B.~R.}\ \bibnamefont {Iyer}},\
  }\href@noop {} {\bibfield  {journal} {\bibinfo  {journal} {\prl}\ }\textbf
  {\bibinfo {volume} {93}},\ \bibinfo {pages} {091101} (\bibinfo {year}
  {2004})},\ \Eprint {http://arxiv.org/abs/0406012} {arXiv:0406012 [gr-qc]}
  \BibitemShut {NoStop}%
\bibitem [{\citenamefont {Damour}\ \emph
  {et~al.}(2001{\natexlab{b}})\citenamefont {Damour}, \citenamefont
  {Jaranowski},\ and\ \citenamefont {Schaefer}}]{Damour:2001bu}%
  \BibitemOpen
  \bibfield  {author} {\bibinfo {author} {\bibfnamefont {T.}~\bibnamefont
  {Damour}}, \bibinfo {author} {\bibfnamefont {P.}~\bibnamefont {Jaranowski}},
  \ and\ \bibinfo {author} {\bibfnamefont {G.}~\bibnamefont {Schaefer}},\
  }\href {\doibase 10.1016/S0370-2693(01)00642-6} {\bibfield  {journal}
  {\bibinfo  {journal} {Phys. Lett. B}\ }\textbf {\bibinfo {volume} {513}},\
  \bibinfo {pages} {147} (\bibinfo {year} {2001}{\natexlab{b}})},\ \Eprint
  {http://arxiv.org/abs/gr-qc/0105038} {arXiv:gr-qc/0105038 [gr-qc]}
  \BibitemShut {NoStop}%
\bibitem [{\citenamefont {Flanagan}(1998)}]{Flanagan:1997fn}%
  \BibitemOpen
  \bibfield  {author} {\bibinfo {author} {\bibfnamefont {E.~E.}\ \bibnamefont
  {Flanagan}},\ }\href {\doibase 10.1103/PhysRevD.58.124030} {\bibfield
  {journal} {\bibinfo  {journal} {\prd}\ }\textbf {\bibinfo {volume} {58}},\
  \bibinfo {pages} {124030} (\bibinfo {year} {1998})},\ \Eprint
  {http://arxiv.org/abs/gr-qc/9706045} {arXiv:gr-qc/9706045 [gr-qc]}
  \BibitemShut {NoStop}%
\bibitem [{\citenamefont {Damour}(1982)}]{Damour:1982wm}%
  \BibitemOpen
  \bibfield  {author} {\bibinfo {author} {\bibfnamefont {T.}~\bibnamefont
  {Damour}},\ }in\ \href@noop {} {\emph {\bibinfo {booktitle} {{Les Houches
  Summer School on Gravitational Radiation Les Houches, France, June 2-21,
  1982}}}}\ (\bibinfo {year} {1982})\BibitemShut {NoStop}%
\bibitem [{\citenamefont {Gralla}(2018)}]{Gralla:2017djj}%
  \BibitemOpen
  \bibfield  {author} {\bibinfo {author} {\bibfnamefont {S.~E.}\ \bibnamefont
  {Gralla}},\ }\href {\doibase 10.1088/1361-6382/aab186} {\bibfield  {journal}
  {\bibinfo  {journal} {Class. Quant. Grav.}\ }\textbf {\bibinfo {volume}
  {35}},\ \bibinfo {pages} {085002} (\bibinfo {year} {2018})},\ \Eprint
  {http://arxiv.org/abs/1710.11096} {arXiv:1710.11096 [gr-qc]} \BibitemShut
  {NoStop}%
\bibitem [{\citenamefont {{Damour}}(1983)}]{Damour:1983a}%
  \BibitemOpen
  \bibfield  {author} {\bibinfo {author} {\bibfnamefont {T.}~\bibnamefont
  {{Damour}}},\ }in\ \href@noop {} {\emph {\bibinfo {booktitle} {Gravitational
  Radiation}}},\ \bibinfo {editor} {edited by\ \bibinfo {editor} {\bibfnamefont
  {N.}~\bibnamefont {{Deruelle}}}\ and\ \bibinfo {editor} {\bibfnamefont
  {T.}~\bibnamefont {{Piran}}}}\ (\bibinfo  {publisher} {North-Holland,
  Amsterdam},\ \bibinfo {year} {1983})\ pp.\ \bibinfo {pages}
  {59--144}\BibitemShut {NoStop}%
\bibitem [{\citenamefont {Hinderer}(2008)}]{Hinderer:2007mb}%
  \BibitemOpen
  \bibfield  {author} {\bibinfo {author} {\bibfnamefont {T.}~\bibnamefont
  {Hinderer}},\ }\href {\doibase 10.1086/533487} {\bibfield  {journal}
  {\bibinfo  {journal} {Astrophys. J.}\ }\textbf {\bibinfo {volume} {677}},\
  \bibinfo {pages} {1216} (\bibinfo {year} {2008})},\ \Eprint
  {http://arxiv.org/abs/0711.2420} {arXiv:0711.2420 [astro-ph]} \BibitemShut
  {NoStop}%
\bibitem [{\citenamefont {{T.~Hinderer, B.~D.~Lackey, R.~N.~Lang and
  J.~S.~Read}}(2010)}]{Hinderer:2009ca}%
  \BibitemOpen
  \bibfield  {author} {\bibinfo {author} {\bibnamefont {{T.~Hinderer,
  B.~D.~Lackey, R.~N.~Lang and J.~S.~Read}}},\ }\href {\doibase
  10.1103/PhysRevD.81.123016} {\bibfield  {journal} {\bibinfo  {journal}
  {\prd}\ }\textbf {\bibinfo {volume} {81}},\ \bibinfo {pages} {123016}
  (\bibinfo {year} {2010})},\ \Eprint {http://arxiv.org/abs/0911.3535}
  {arXiv:0911.3535 [astro-ph]} \BibitemShut {NoStop}%
\bibitem [{\citenamefont {Binnington}\ and\ \citenamefont
  {Poisson}(2009)}]{Binnington:2009bb}%
  \BibitemOpen
  \bibfield  {author} {\bibinfo {author} {\bibfnamefont {T.}~\bibnamefont
  {Binnington}}\ and\ \bibinfo {author} {\bibfnamefont {E.}~\bibnamefont
  {Poisson}},\ }\href {\doibase 10.1103/PhysRevD.80.084018} {\bibfield
  {journal} {\bibinfo  {journal} {\prd}\ }\textbf {\bibinfo {volume} {80}},\
  \bibinfo {pages} {084018} (\bibinfo {year} {2009})},\ \Eprint
  {http://arxiv.org/abs/0906.1366} {arXiv:0906.1366 [gr-qc]} \BibitemShut
  {NoStop}%
\bibitem [{\citenamefont {Damour}\ and\ \citenamefont
  {Nagar}(2009)}]{Damour:2009vw}%
  \BibitemOpen
  \bibfield  {author} {\bibinfo {author} {\bibfnamefont {T.}~\bibnamefont
  {Damour}}\ and\ \bibinfo {author} {\bibfnamefont {A.}~\bibnamefont {Nagar}},\
  }\href {\doibase 10.1103/PhysRevD.80.084035} {\bibfield  {journal} {\bibinfo
  {journal} {\prd}\ }\textbf {\bibinfo {volume} {80}},\ \bibinfo {pages}
  {084035} (\bibinfo {year} {2009})},\ \Eprint {http://arxiv.org/abs/0906.0096}
  {arXiv:0906.0096 [gr-qc]} \BibitemShut {NoStop}%
\bibitem [{\citenamefont {Poisson}(1998)}]{Poisson:1997ha}%
  \BibitemOpen
  \bibfield  {author} {\bibinfo {author} {\bibfnamefont {E.}~\bibnamefont
  {Poisson}},\ }\href {\doibase 10.1103/PhysRevD.57.5287} {\bibfield  {journal}
  {\bibinfo  {journal} {\prd}\ }\textbf {\bibinfo {volume} {57}},\ \bibinfo
  {pages} {5287} (\bibinfo {year} {1998})},\ \Eprint
  {http://arxiv.org/abs/gr-qc/9709032} {arXiv:gr-qc/9709032 [gr-qc]}
  \BibitemShut {NoStop}%
\bibitem [{\citenamefont {Boh{\'e}}\ \emph {et~al.}(2015)\citenamefont
  {Boh{\'e}}, \citenamefont {Faye}, \citenamefont {Marsat},\ and\ \citenamefont
  {Porter}}]{Bohe:2015ana}%
  \BibitemOpen
  \bibfield  {author} {\bibinfo {author} {\bibfnamefont {A.}~\bibnamefont
  {Boh{\'e}}}, \bibinfo {author} {\bibfnamefont {G.}~\bibnamefont {Faye}},
  \bibinfo {author} {\bibfnamefont {S.}~\bibnamefont {Marsat}}, \ and\ \bibinfo
  {author} {\bibfnamefont {E.~K.}\ \bibnamefont {Porter}},\ }\href {\doibase
  10.1088/0264-9381/32/19/195010} {\bibfield  {journal} {\bibinfo  {journal}
  {Class. Quant. Grav.}\ }\textbf {\bibinfo {volume} {32}},\ \bibinfo {pages}
  {195010} (\bibinfo {year} {2015})},\ \Eprint
  {http://arxiv.org/abs/1501.01529} {arXiv:1501.01529 [gr-qc]} \BibitemShut
  {NoStop}%
\bibitem [{\citenamefont {Baiotti}\ and\ \citenamefont
  {Rezzolla}(2017)}]{Baiotti:2016qnr}%
  \BibitemOpen
  \bibfield  {author} {\bibinfo {author} {\bibfnamefont {L.}~\bibnamefont
  {Baiotti}}\ and\ \bibinfo {author} {\bibfnamefont {L.}~\bibnamefont
  {Rezzolla}},\ }\href {\doibase 10.1088/1361-6633/aa67bb} {\bibfield
  {journal} {\bibinfo  {journal} {Rept. Prog. Phys.}\ }\textbf {\bibinfo
  {volume} {80}},\ \bibinfo {pages} {096901} (\bibinfo {year} {2017})},\
  \Eprint {http://arxiv.org/abs/1607.03540} {arXiv:1607.03540 [gr-qc]}
  \BibitemShut {NoStop}%
\bibitem [{\citenamefont {{Lai}}(1994)}]{1994MNRAS.270..611L}%
  \BibitemOpen
  \bibfield  {author} {\bibinfo {author} {\bibfnamefont {D.}~\bibnamefont
  {{Lai}}},\ }\href {\doibase 10.1093/mnras/270.3.611} {\bibfield  {journal}
  {\bibinfo  {journal} {Monthly Notices of the Royal Astronomical Society}\
  }\textbf {\bibinfo {volume} {270}},\ \bibinfo {pages} {611} (\bibinfo {year}
  {1994})},\ \Eprint {http://arxiv.org/abs/astro-ph/9404062} {astro-ph/9404062}
  \BibitemShut {NoStop}%
\bibitem [{\citenamefont {{Hinderer}}\ \emph {et~al.}(2016)\citenamefont
  {{Hinderer}}, \citenamefont {{Taracchini}}, \citenamefont {{Foucart}},
  \citenamefont {{Buonanno}}, \citenamefont {{Steinhoff}}, \citenamefont
  {{Duez}}, \citenamefont {{Kidder}}, \citenamefont {{Pfeiffer}}, \citenamefont
  {{Scheel}}, \citenamefont {{Szilagyi}}, \citenamefont {{Hotokezaka}},
  \citenamefont {{Kyutoku}}, \citenamefont {{Shibata}},\ and\ \citenamefont
  {{Carpenter}}}]{2016PhRvL.116r1101H}%
  \BibitemOpen
  \bibfield  {author} {\bibinfo {author} {\bibfnamefont {T.}~\bibnamefont
  {{Hinderer}}}, \bibinfo {author} {\bibfnamefont {A.}~\bibnamefont
  {{Taracchini}}}, \bibinfo {author} {\bibfnamefont {F.}~\bibnamefont
  {{Foucart}}}, \bibinfo {author} {\bibfnamefont {A.}~\bibnamefont
  {{Buonanno}}}, \bibinfo {author} {\bibfnamefont {J.}~\bibnamefont
  {{Steinhoff}}}, \bibinfo {author} {\bibfnamefont {M.}~\bibnamefont {{Duez}}},
  \bibinfo {author} {\bibfnamefont {L.~E.}\ \bibnamefont {{Kidder}}}, \bibinfo
  {author} {\bibfnamefont {H.~P.}\ \bibnamefont {{Pfeiffer}}}, \bibinfo
  {author} {\bibfnamefont {M.~A.}\ \bibnamefont {{Scheel}}}, \bibinfo {author}
  {\bibfnamefont {B.}~\bibnamefont {{Szilagyi}}}, \bibinfo {author}
  {\bibfnamefont {K.}~\bibnamefont {{Hotokezaka}}}, \bibinfo {author}
  {\bibfnamefont {K.}~\bibnamefont {{Kyutoku}}}, \bibinfo {author}
  {\bibfnamefont {M.}~\bibnamefont {{Shibata}}}, \ and\ \bibinfo {author}
  {\bibfnamefont {C.~W.}\ \bibnamefont {{Carpenter}}},\ }\href {\doibase
  10.1103/PhysRevLett.116.181101} {\bibfield  {journal} {\bibinfo  {journal}
  {Physical Review Letters}\ }\textbf {\bibinfo {volume} {116}},\ \bibinfo
  {eid} {181101} (\bibinfo {year} {2016})},\ \Eprint
  {http://arxiv.org/abs/1602.00599} {arXiv:1602.00599 [gr-qc]} \BibitemShut
  {NoStop}%
\bibitem [{\citenamefont {{Andersson}}\ and\ \citenamefont
  {{Ho}}(2018)}]{2018PhRvD..97b3016A}%
  \BibitemOpen
  \bibfield  {author} {\bibinfo {author} {\bibfnamefont {N.}~\bibnamefont
  {{Andersson}}}\ and\ \bibinfo {author} {\bibfnamefont {W.~C.~G.}\
  \bibnamefont {{Ho}}},\ }\href {\doibase 10.1103/PhysRevD.97.023016}
  {\bibfield  {journal} {\bibinfo  {journal} {\prd}\ }\textbf {\bibinfo
  {volume} {97}},\ \bibinfo {eid} {023016} (\bibinfo {year} {2018})},\ \Eprint
  {http://arxiv.org/abs/1710.05950} {arXiv:1710.05950 [astro-ph.HE]}
  \BibitemShut {NoStop}%
\bibitem [{\citenamefont {Laarakkers}\ and\ \citenamefont
  {Poisson}(1999)}]{Laarakkers:1997hb}%
  \BibitemOpen
  \bibfield  {author} {\bibinfo {author} {\bibfnamefont {W.~G.}\ \bibnamefont
  {Laarakkers}}\ and\ \bibinfo {author} {\bibfnamefont {E.}~\bibnamefont
  {Poisson}},\ }\href {\doibase 10.1086/306732} {\bibfield  {journal} {\bibinfo
   {journal} {Astrophys. J.}\ }\textbf {\bibinfo {volume} {512}},\ \bibinfo
  {pages} {282} (\bibinfo {year} {1999})},\ \Eprint
  {http://arxiv.org/abs/gr-qc/9709033} {arXiv:gr-qc/9709033 [gr-qc]}
  \BibitemShut {NoStop}%
\bibitem [{\citenamefont {Pappas}\ and\ \citenamefont
  {Apostolatos}(2012)}]{Pappas:2012qg}%
  \BibitemOpen
  \bibfield  {author} {\bibinfo {author} {\bibfnamefont {G.}~\bibnamefont
  {Pappas}}\ and\ \bibinfo {author} {\bibfnamefont {T.~A.}\ \bibnamefont
  {Apostolatos}},\ }\href@noop {} {\  (\bibinfo {year} {2012})},\ \Eprint
  {http://arxiv.org/abs/1211.6299} {arXiv:1211.6299 [gr-qc]} \BibitemShut
  {NoStop}%
\bibitem [{\citenamefont {Agathos}\ \emph {et~al.}(2015)\citenamefont
  {Agathos}, \citenamefont {Meidam}, \citenamefont {Del~Pozzo}, \citenamefont
  {Li}, \citenamefont {Tompitak}, \citenamefont {Veitch}, \citenamefont
  {Vitale},\ and\ \citenamefont {Van Den~Broeck}}]{PhysRevD.92.023012}%
  \BibitemOpen
  \bibfield  {author} {\bibinfo {author} {\bibfnamefont {M.}~\bibnamefont
  {Agathos}}, \bibinfo {author} {\bibfnamefont {J.}~\bibnamefont {Meidam}},
  \bibinfo {author} {\bibfnamefont {W.}~\bibnamefont {Del~Pozzo}}, \bibinfo
  {author} {\bibfnamefont {T.~G.~F.}\ \bibnamefont {Li}}, \bibinfo {author}
  {\bibfnamefont {M.}~\bibnamefont {Tompitak}}, \bibinfo {author}
  {\bibfnamefont {J.}~\bibnamefont {Veitch}}, \bibinfo {author} {\bibfnamefont
  {S.}~\bibnamefont {Vitale}}, \ and\ \bibinfo {author} {\bibfnamefont
  {C.}~\bibnamefont {Van Den~Broeck}},\ }\href {\doibase
  10.1103/PhysRevD.92.023012} {\bibfield  {journal} {\bibinfo  {journal} {Phys.
  Rev. D}\ }\textbf {\bibinfo {volume} {92}},\ \bibinfo {pages} {023012}
  (\bibinfo {year} {2015})}\BibitemShut {NoStop}%
\bibitem [{\citenamefont {Harry}\ and\ \citenamefont
  {Hinderer}(2018)}]{Harry:2018hke}%
  \BibitemOpen
  \bibfield  {author} {\bibinfo {author} {\bibfnamefont {I.}~\bibnamefont
  {Harry}}\ and\ \bibinfo {author} {\bibfnamefont {T.}~\bibnamefont
  {Hinderer}},\ }\href@noop {} {\  (\bibinfo {year} {2018})},\ \Eprint
  {http://arxiv.org/abs/1801.09972} {arXiv:1801.09972 [gr-qc]} \BibitemShut
  {NoStop}%
\bibitem [{\citenamefont {Krishnendu}\ \emph {et~al.}(2017)\citenamefont
  {Krishnendu}, \citenamefont {Arun},\ and\ \citenamefont
  {Mishra}}]{Krishnendu:2017shb}%
  \BibitemOpen
  \bibfield  {author} {\bibinfo {author} {\bibfnamefont {N.~V.}\ \bibnamefont
  {Krishnendu}}, \bibinfo {author} {\bibfnamefont {K.~G.}\ \bibnamefont
  {Arun}}, \ and\ \bibinfo {author} {\bibfnamefont {C.~K.}\ \bibnamefont
  {Mishra}},\ }\href {\doibase 10.1103/PhysRevLett.119.091101} {\bibfield
  {journal} {\bibinfo  {journal} {Phys. Rev. Lett.}\ }\textbf {\bibinfo
  {volume} {119}},\ \bibinfo {pages} {091101} (\bibinfo {year} {2017})},\
  \Eprint {http://arxiv.org/abs/1701.06318} {arXiv:1701.06318 [gr-qc]}
  \BibitemShut {NoStop}%
\bibitem [{\citenamefont {{Abbott}}\ \emph
  {et~al.}(2017{\natexlab{b}})\citenamefont {{Abbott}} \emph
  {et~al.}}]{PostmergerSearchPaper}%
  \BibitemOpen
  \bibfield  {author} {\bibinfo {author} {\bibfnamefont {B.~P.}\ \bibnamefont
  {{Abbott}}} \emph {et~al.} (\bibinfo {collaboration} {LIGO Scientific
  Collaboration, Virgo Collaboration}),\ }\href {\doibase
  10.3847/2041-8213/aa9a35} {\bibfield  {journal} {\bibinfo  {journal}
  {Astrophys. J.}\ }\textbf {\bibinfo {volume} {851}},\ \bibinfo {pages} {L16}
  (\bibinfo {year} {2017}{\natexlab{b}})},\ \Eprint
  {http://arxiv.org/abs/1710.09320} {arXiv:1710.09320 [astro-ph.HE]}
  \BibitemShut {NoStop}%
\bibitem [{\citenamefont {Abbott}\ \emph
  {et~al.}(2018{\natexlab{b}})\citenamefont {Abbott} \emph
  {et~al.}}]{Abbott:2018wiz}%
  \BibitemOpen
  \bibfield  {author} {\bibinfo {author} {\bibfnamefont {B.~P.}\ \bibnamefont
  {Abbott}} \emph {et~al.} (\bibinfo {collaboration} {Virgo, LIGO
  Scientific}),\ }\href@noop {} {\  (\bibinfo {year} {2018}{\natexlab{b}})},\
  \bibinfo {note} {{}LIGO-P1800061,
  \url{https://dcc.ligo.org/LIGO-P1800061/public}},\ \Eprint
  {http://arxiv.org/abs/1805.11579} {arXiv:1805.11579 [gr-qc]} \BibitemShut
  {NoStop}%
\bibitem [{\citenamefont {De}\ \emph {et~al.}(2018)\citenamefont {De},
  \citenamefont {Finstad}, \citenamefont {Lattimer}, \citenamefont {Brown},
  \citenamefont {Berger},\ and\ \citenamefont {Biwer}}]{De:2018uhw}%
  \BibitemOpen
  \bibfield  {author} {\bibinfo {author} {\bibfnamefont {S.}~\bibnamefont
  {De}}, \bibinfo {author} {\bibfnamefont {D.}~\bibnamefont {Finstad}},
  \bibinfo {author} {\bibfnamefont {J.~M.}\ \bibnamefont {Lattimer}}, \bibinfo
  {author} {\bibfnamefont {D.~A.}\ \bibnamefont {Brown}}, \bibinfo {author}
  {\bibfnamefont {E.}~\bibnamefont {Berger}}, \ and\ \bibinfo {author}
  {\bibfnamefont {C.~M.}\ \bibnamefont {Biwer}},\ }\href {\doibase
  10.1103/PhysRevLett.121.091102} {\bibfield  {journal} {\bibinfo  {journal}
  {Phys. Rev. Lett.}\ }\textbf {\bibinfo {volume} {121}},\ \bibinfo {pages}
  {091102} (\bibinfo {year} {2018})},\ \Eprint
  {http://arxiv.org/abs/1804.08583} {arXiv:1804.08583 [astro-ph.HE]}
  \BibitemShut {NoStop}%
\bibitem [{\citenamefont {Margalit}\ and\ \citenamefont
  {Metzger}(2017)}]{2041-8205-850-2-L19}%
  \BibitemOpen
  \bibfield  {author} {\bibinfo {author} {\bibfnamefont {B.}~\bibnamefont
  {Margalit}}\ and\ \bibinfo {author} {\bibfnamefont {B.~D.}\ \bibnamefont
  {Metzger}},\ }\href {http://stacks.iop.org/2041-8205/850/i=2/a=L19}
  {\bibfield  {journal} {\bibinfo  {journal} {The Astrophysical Journal
  Letters}\ }\textbf {\bibinfo {volume} {850}},\ \bibinfo {pages} {L19}
  (\bibinfo {year} {2017})}\BibitemShut {NoStop}%
\bibitem [{\citenamefont {Bauswein}\ \emph {et~al.}(2017)\citenamefont
  {Bauswein}, \citenamefont {Just}, \citenamefont {Janka},\ and\ \citenamefont
  {Stergioulas}}]{2041-8205-850-2-L34}%
  \BibitemOpen
  \bibfield  {author} {\bibinfo {author} {\bibfnamefont {A.}~\bibnamefont
  {Bauswein}}, \bibinfo {author} {\bibfnamefont {O.}~\bibnamefont {Just}},
  \bibinfo {author} {\bibfnamefont {H.-T.}\ \bibnamefont {Janka}}, \ and\
  \bibinfo {author} {\bibfnamefont {N.}~\bibnamefont {Stergioulas}},\ }\href
  {http://stacks.iop.org/2041-8205/850/i=2/a=L34} {\bibfield  {journal}
  {\bibinfo  {journal} {The Astrophysical Journal Letters}\ }\textbf {\bibinfo
  {volume} {850}},\ \bibinfo {pages} {L34} (\bibinfo {year}
  {2017})}\BibitemShut {NoStop}%
\bibitem [{\citenamefont {Zhou}\ \emph {et~al.}(2018)\citenamefont {Zhou},
  \citenamefont {Zhou},\ and\ \citenamefont {Li}}]{Zhou:2017pha}%
  \BibitemOpen
  \bibfield  {author} {\bibinfo {author} {\bibfnamefont {E.-P.}\ \bibnamefont
  {Zhou}}, \bibinfo {author} {\bibfnamefont {X.}~\bibnamefont {Zhou}}, \ and\
  \bibinfo {author} {\bibfnamefont {A.}~\bibnamefont {Li}},\ }\href {\doibase
  10.1103/PhysRevD.97.083015} {\bibfield  {journal} {\bibinfo  {journal}
  {\prd}\ }\textbf {\bibinfo {volume} {97}},\ \bibinfo {pages} {083015}
  (\bibinfo {year} {2018})},\ \Eprint {http://arxiv.org/abs/1711.04312}
  {arXiv:1711.04312 [astro-ph.HE]} \BibitemShut {NoStop}%
\bibitem [{\citenamefont {Rezzolla}\ \emph {et~al.}(2018)\citenamefont
  {Rezzolla}, \citenamefont {Most},\ and\ \citenamefont
  {Weih}}]{2041-8205-852-2-L25}%
  \BibitemOpen
  \bibfield  {author} {\bibinfo {author} {\bibfnamefont {L.}~\bibnamefont
  {Rezzolla}}, \bibinfo {author} {\bibfnamefont {E.~R.}\ \bibnamefont {Most}},
  \ and\ \bibinfo {author} {\bibfnamefont {L.~R.}\ \bibnamefont {Weih}},\
  }\href {http://stacks.iop.org/2041-8205/852/i=2/a=L25} {\bibfield  {journal}
  {\bibinfo  {journal} {The Astrophysical Journal Letters}\ }\textbf {\bibinfo
  {volume} {852}},\ \bibinfo {pages} {L25} (\bibinfo {year}
  {2018})}\BibitemShut {NoStop}%
\bibitem [{\citenamefont {Fattoyev}\ \emph {et~al.}(2018)\citenamefont
  {Fattoyev}, \citenamefont {Piekarewicz},\ and\ \citenamefont
  {Horowitz}}]{PhysRevLett.120.172702}%
  \BibitemOpen
  \bibfield  {author} {\bibinfo {author} {\bibfnamefont {F.~J.}\ \bibnamefont
  {Fattoyev}}, \bibinfo {author} {\bibfnamefont {J.}~\bibnamefont
  {Piekarewicz}}, \ and\ \bibinfo {author} {\bibfnamefont {C.~J.}\ \bibnamefont
  {Horowitz}},\ }\href {\doibase 10.1103/PhysRevLett.120.172702} {\bibfield
  {journal} {\bibinfo  {journal} {Phys. Rev. Lett.}\ }\textbf {\bibinfo
  {volume} {120}},\ \bibinfo {pages} {172702} (\bibinfo {year}
  {2018})}\BibitemShut {NoStop}%
\bibitem [{\citenamefont {Nandi}\ and\ \citenamefont
  {Char}(2018)}]{Nandi:2017rhy}%
  \BibitemOpen
  \bibfield  {author} {\bibinfo {author} {\bibfnamefont {R.}~\bibnamefont
  {Nandi}}\ and\ \bibinfo {author} {\bibfnamefont {P.}~\bibnamefont {Char}},\
  }\href {\doibase 10.3847/1538-4357/aab78c} {\bibfield  {journal} {\bibinfo
  {journal} {Astrophys. J.}\ }\textbf {\bibinfo {volume} {857}},\ \bibinfo
  {pages} {12} (\bibinfo {year} {2018})},\ \Eprint
  {http://arxiv.org/abs/1712.08094} {arXiv:1712.08094 [astro-ph.HE]}
  \BibitemShut {NoStop}%
\bibitem [{\citenamefont {Paschalidis}\ \emph {et~al.}(2018)\citenamefont
  {Paschalidis}, \citenamefont {Yagi}, \citenamefont {Alvarez-Castillo},
  \citenamefont {Blaschke},\ and\ \citenamefont
  {Sedrakian}}]{Paschalidis:2017qmb}%
  \BibitemOpen
  \bibfield  {author} {\bibinfo {author} {\bibfnamefont {V.}~\bibnamefont
  {Paschalidis}}, \bibinfo {author} {\bibfnamefont {K.}~\bibnamefont {Yagi}},
  \bibinfo {author} {\bibfnamefont {D.}~\bibnamefont {Alvarez-Castillo}},
  \bibinfo {author} {\bibfnamefont {D.~B.}\ \bibnamefont {Blaschke}}, \ and\
  \bibinfo {author} {\bibfnamefont {A.}~\bibnamefont {Sedrakian}},\ }\href
  {\doibase 10.1103/PhysRevD.97.084038} {\bibfield  {journal} {\bibinfo
  {journal} {\prd}\ }\textbf {\bibinfo {volume} {97}},\ \bibinfo {pages}
  {084038} (\bibinfo {year} {2018})},\ \Eprint
  {http://arxiv.org/abs/1712.00451} {arXiv:1712.00451 [astro-ph.HE]}
  \BibitemShut {NoStop}%
\bibitem [{\citenamefont {Ruiz}\ \emph {et~al.}(2018)\citenamefont {Ruiz},
  \citenamefont {Shapiro},\ and\ \citenamefont
  {Tsokaros}}]{PhysRevD.97.021501}%
  \BibitemOpen
  \bibfield  {author} {\bibinfo {author} {\bibfnamefont {M.}~\bibnamefont
  {Ruiz}}, \bibinfo {author} {\bibfnamefont {S.~L.}\ \bibnamefont {Shapiro}}, \
  and\ \bibinfo {author} {\bibfnamefont {A.}~\bibnamefont {Tsokaros}},\ }\href
  {\doibase 10.1103/PhysRevD.97.021501} {\bibfield  {journal} {\bibinfo
  {journal} {Phys. Rev. D}\ }\textbf {\bibinfo {volume} {97}},\ \bibinfo
  {pages} {021501} (\bibinfo {year} {2018})}\BibitemShut {NoStop}%
\bibitem [{\citenamefont {Annala}\ \emph {et~al.}(2018)\citenamefont {Annala},
  \citenamefont {Gorda}, \citenamefont {Kurkela},\ and\ \citenamefont
  {Vuorinen}}]{PhysRevLett.120.172703}%
  \BibitemOpen
  \bibfield  {author} {\bibinfo {author} {\bibfnamefont {E.}~\bibnamefont
  {Annala}}, \bibinfo {author} {\bibfnamefont {T.}~\bibnamefont {Gorda}},
  \bibinfo {author} {\bibfnamefont {A.}~\bibnamefont {Kurkela}}, \ and\
  \bibinfo {author} {\bibfnamefont {A.}~\bibnamefont {Vuorinen}},\ }\href
  {\doibase 10.1103/PhysRevLett.120.172703} {\bibfield  {journal} {\bibinfo
  {journal} {Phys. Rev. Lett.}\ }\textbf {\bibinfo {volume} {120}},\ \bibinfo
  {pages} {172703} (\bibinfo {year} {2018})}\BibitemShut {NoStop}%
\bibitem [{\citenamefont {Raithel}\ \emph {et~al.}(2018)\citenamefont
  {Raithel}, \citenamefont {{\"{O}}zel},\ and\ \citenamefont
  {Psaltis}}]{Raithel:2018ncd}%
  \BibitemOpen
  \bibfield  {author} {\bibinfo {author} {\bibfnamefont {C.}~\bibnamefont
  {Raithel}}, \bibinfo {author} {\bibfnamefont {F.}~\bibnamefont {{\"{O}}zel}},
  \ and\ \bibinfo {author} {\bibfnamefont {D.}~\bibnamefont {Psaltis}},\
  }\href@noop {} {\  (\bibinfo {year} {2018})},\ \Eprint
  {http://arxiv.org/abs/1803.07687} {arXiv:1803.07687 [astro-ph.HE]}
  \BibitemShut {NoStop}%
\bibitem [{\citenamefont {Most}\ \emph {et~al.}(2018)\citenamefont {Most},
  \citenamefont {Weih}, \citenamefont {Rezzolla},\ and\ \citenamefont
  {Schaffner-Bielich}}]{Most:2018hfd}%
  \BibitemOpen
  \bibfield  {author} {\bibinfo {author} {\bibfnamefont {E.~R.}\ \bibnamefont
  {Most}}, \bibinfo {author} {\bibfnamefont {L.~R.}\ \bibnamefont {Weih}},
  \bibinfo {author} {\bibfnamefont {L.}~\bibnamefont {Rezzolla}}, \ and\
  \bibinfo {author} {\bibfnamefont {J.}~\bibnamefont {Schaffner-Bielich}},\
  }\href@noop {} {\  (\bibinfo {year} {2018})},\ \Eprint
  {http://arxiv.org/abs/1803.00549} {arXiv:1803.00549 [gr-qc]} \BibitemShut
  {NoStop}%
\bibitem [{\citenamefont {Antoniadis}\ \emph {et~al.}(2013)\citenamefont
  {Antoniadis}, \citenamefont {Freire}, \citenamefont {Wex}, \citenamefont
  {Tauris}, \citenamefont {Lynch} \emph {et~al.}}]{Antoniadis:2013pzd}%
  \BibitemOpen
  \bibfield  {author} {\bibinfo {author} {\bibfnamefont {J.}~\bibnamefont
  {Antoniadis}}, \bibinfo {author} {\bibfnamefont {P.~C.}\ \bibnamefont
  {Freire}}, \bibinfo {author} {\bibfnamefont {N.}~\bibnamefont {Wex}},
  \bibinfo {author} {\bibfnamefont {T.~M.}\ \bibnamefont {Tauris}}, \bibinfo
  {author} {\bibfnamefont {R.~S.}\ \bibnamefont {Lynch}},  \emph {et~al.},\
  }\href {\doibase 10.1126/science.1233232} {\bibfield  {journal} {\bibinfo
  {journal} {Science}\ }\textbf {\bibinfo {volume} {340}},\ \bibinfo {pages}
  {1233232} (\bibinfo {year} {2013})},\ \Eprint
  {http://arxiv.org/abs/1304.6875} {arXiv:1304.6875 [astro-ph.HE]} \BibitemShut
  {NoStop}%
\bibitem [{\citenamefont {{Gravitational Wave Open Science Center
  (GWOSC)}}()}]{GWOSC}%
  \BibitemOpen
  \bibfield  {author} {\bibinfo {author} {\bibnamefont {{Gravitational Wave
  Open Science Center (GWOSC)}}},\ }\href@noop {} {}\bibinfo {howpublished}
  {\url{https://www.gw-openscience.org}}\BibitemShut {NoStop}%
\bibitem [{\citenamefont {Pankow}\ \emph {et~al.}(2018)\citenamefont {Pankow}
  \emph {et~al.}}]{Pankow:2018qpo}%
  \BibitemOpen
  \bibfield  {author} {\bibinfo {author} {\bibfnamefont {C.}~\bibnamefont
  {Pankow}} \emph {et~al.},\ }\href@noop {} {\  (\bibinfo {year} {2018})},\
  \Eprint {http://arxiv.org/abs/1808.03619} {arXiv:1808.03619 [gr-qc]}
  \BibitemShut {NoStop}%
\bibitem [{\citenamefont {Driggers}\ \emph {et~al.}(2018)\citenamefont
  {Driggers} \emph {et~al.}}]{Driggers:2018gii}%
  \BibitemOpen
  \bibfield  {author} {\bibinfo {author} {\bibfnamefont {J.~C.}\ \bibnamefont
  {Driggers}} \emph {et~al.} (\bibinfo {collaboration} {LIGO Scientific}),\
  }\href {https://dcc.ligo.org/LIGO-P1700260/public} {\  (\bibinfo {year}
  {2018})},\ \Eprint {http://arxiv.org/abs/1806.00532} {arXiv:1806.00532
  [astro-ph.IM]} \BibitemShut {NoStop}%
\bibitem [{\citenamefont {Driggers}\ \emph {et~al.}(2012)\citenamefont
  {Driggers}, \citenamefont {Evans}, \citenamefont {Pepper},\ and\
  \citenamefont {Adhikari}}]{iLIGO_SeisCleaning}%
  \BibitemOpen
  \bibfield  {author} {\bibinfo {author} {\bibfnamefont {J.~C.}\ \bibnamefont
  {Driggers}}, \bibinfo {author} {\bibfnamefont {M.}~\bibnamefont {Evans}},
  \bibinfo {author} {\bibfnamefont {K.}~\bibnamefont {Pepper}}, \ and\ \bibinfo
  {author} {\bibfnamefont {R.}~\bibnamefont {Adhikari}},\ }\href {\doibase
  10.1063/1.3675891} {\bibfield  {journal} {\bibinfo  {journal}
  {Rev.~Sci.~Instrum.}\ }\textbf {\bibinfo {volume} {83}},\ \bibinfo {pages}
  {024501} (\bibinfo {year} {2012})},\ \Eprint {http://arxiv.org/abs/1112.2224}
  {arXiv:1112.2224 [gr-qc]} \BibitemShut {NoStop}%
\bibitem [{\citenamefont {Meadors}\ \emph {et~al.}(2014)\citenamefont
  {Meadors}, \citenamefont {Kawabe},\ and\ \citenamefont
  {Riles}}]{iLIGO_LSCCleaning}%
  \BibitemOpen
  \bibfield  {author} {\bibinfo {author} {\bibfnamefont {G.~D.}\ \bibnamefont
  {Meadors}}, \bibinfo {author} {\bibfnamefont {K.}~\bibnamefont {Kawabe}}, \
  and\ \bibinfo {author} {\bibfnamefont {K.}~\bibnamefont {Riles}},\ }\href
  {\doibase 10.1088/0264-9381/31/10/105014} {\bibfield  {journal} {\bibinfo
  {journal} {Class.~Quantum.~Grav.}\ }\textbf {\bibinfo {volume} {31}},\
  \bibinfo {pages} {105014} (\bibinfo {year} {2014})},\ \Eprint
  {http://arxiv.org/abs/1311.6835} {arXiv:1311.6835 [astro-ph.IM]} \BibitemShut
  {NoStop}%
\bibitem [{\citenamefont {Tiwari}\ \emph {et~al.}(2015)\citenamefont {Tiwari}
  \emph {et~al.}}]{iLIGO_PEMCleaning}%
  \BibitemOpen
  \bibfield  {author} {\bibinfo {author} {\bibfnamefont {V.}~\bibnamefont
  {Tiwari}} \emph {et~al.},\ }\href {\doibase 10.1088/0264-9381/32/16/165014}
  {\bibfield  {journal} {\bibinfo  {journal} {Class.~Quantum.~Grav.}\ }\textbf
  {\bibinfo {volume} {32}},\ \bibinfo {pages} {165014} (\bibinfo {year}
  {2015})},\ \Eprint {http://arxiv.org/abs/1503.07476} {arXiv:1503.07476
  [gr-qc]} \BibitemShut {NoStop}%
\bibitem [{\citenamefont {Veitch}\ \emph {et~al.}(2015)\citenamefont {Veitch},
  \citenamefont {Raymond}, \citenamefont {Farr}, \citenamefont {Farr},
  \citenamefont {Graff}, \citenamefont {Vitale}, \citenamefont {Aylott},
  \citenamefont {Blackburn}, \citenamefont {Christensen}, \citenamefont
  {Coughlin}, \citenamefont {Del~Pozzo}, \citenamefont {Feroz}, \citenamefont
  {Gair}, \citenamefont {Haster}, \citenamefont {Kalogera}, \citenamefont
  {Littenberg}, \citenamefont {Mandel}, \citenamefont {O'Shaughnessy},
  \citenamefont {Pitkin}, \citenamefont {Rodriguez}, \citenamefont {R\"over},
  \citenamefont {Sidery}, \citenamefont {Smith}, \citenamefont {Van Der~Sluys},
  \citenamefont {Vecchio}, \citenamefont {Vousden},\ and\ \citenamefont
  {Wade}}]{Veitch:2014wba}%
  \BibitemOpen
  \bibfield  {author} {\bibinfo {author} {\bibfnamefont {J.}~\bibnamefont
  {Veitch}}, \bibinfo {author} {\bibfnamefont {V.}~\bibnamefont {Raymond}},
  \bibinfo {author} {\bibfnamefont {B.}~\bibnamefont {Farr}}, \bibinfo {author}
  {\bibfnamefont {W.}~\bibnamefont {Farr}}, \bibinfo {author} {\bibfnamefont
  {P.}~\bibnamefont {Graff}}, \bibinfo {author} {\bibfnamefont
  {S.}~\bibnamefont {Vitale}}, \bibinfo {author} {\bibfnamefont
  {B.}~\bibnamefont {Aylott}}, \bibinfo {author} {\bibfnamefont
  {K.}~\bibnamefont {Blackburn}}, \bibinfo {author} {\bibfnamefont
  {N.}~\bibnamefont {Christensen}}, \bibinfo {author} {\bibfnamefont
  {M.}~\bibnamefont {Coughlin}}, \bibinfo {author} {\bibfnamefont
  {W.}~\bibnamefont {Del~Pozzo}}, \bibinfo {author} {\bibfnamefont
  {F.}~\bibnamefont {Feroz}}, \bibinfo {author} {\bibfnamefont
  {J.}~\bibnamefont {Gair}}, \bibinfo {author} {\bibfnamefont {C.-J.}\
  \bibnamefont {Haster}}, \bibinfo {author} {\bibfnamefont {V.}~\bibnamefont
  {Kalogera}}, \bibinfo {author} {\bibfnamefont {T.}~\bibnamefont
  {Littenberg}}, \bibinfo {author} {\bibfnamefont {I.}~\bibnamefont {Mandel}},
  \bibinfo {author} {\bibfnamefont {R.}~\bibnamefont {O'Shaughnessy}}, \bibinfo
  {author} {\bibfnamefont {M.}~\bibnamefont {Pitkin}}, \bibinfo {author}
  {\bibfnamefont {C.}~\bibnamefont {Rodriguez}}, \bibinfo {author}
  {\bibfnamefont {C.}~\bibnamefont {R\"over}}, \bibinfo {author} {\bibfnamefont
  {T.}~\bibnamefont {Sidery}}, \bibinfo {author} {\bibfnamefont
  {R.}~\bibnamefont {Smith}}, \bibinfo {author} {\bibfnamefont
  {M.}~\bibnamefont {Van Der~Sluys}}, \bibinfo {author} {\bibfnamefont
  {A.}~\bibnamefont {Vecchio}}, \bibinfo {author} {\bibfnamefont
  {W.}~\bibnamefont {Vousden}}, \ and\ \bibinfo {author} {\bibfnamefont
  {L.}~\bibnamefont {Wade}},\ }\href {\doibase 10.1103/PhysRevD.91.042003}
  {\bibfield  {journal} {\bibinfo  {journal} {Phys. Rev. D}\ }\textbf {\bibinfo
  {volume} {91}},\ \bibinfo {pages} {042003} (\bibinfo {year} {2015})},\
  \Eprint {http://arxiv.org/abs/1409.7215} {arXiv:1409.7215 [gr-qc]}
  \BibitemShut {NoStop}%
\bibitem [{\citenamefont {Abbott}\ \emph {et~al.}(2016)\citenamefont {Abbott}
  \emph {et~al.}}]{TheLIGOScientific:2016wfe}%
  \BibitemOpen
  \bibfield  {author} {\bibinfo {author} {\bibfnamefont {B.~P.}\ \bibnamefont
  {Abbott}} \emph {et~al.} (\bibinfo {collaboration} {LIGO Scientific
  Collaboration, Virgo Collaboration}),\ }\href {\doibase
  10.1103/PhysRevLett.116.241102} {\bibfield  {journal} {\bibinfo  {journal}
  {Phys. Rev. Lett.}\ }\textbf {\bibinfo {volume} {116}},\ \bibinfo {pages}
  {241102} (\bibinfo {year} {2016})},\ \Eprint
  {http://arxiv.org/abs/1602.03840} {arXiv:1602.03840 [gr-qc]} \BibitemShut
  {NoStop}%
\bibitem [{\citenamefont {{LIGO Scientific Collaboration and Virgo
  Collaboration}}(2017)}]{lalinference_o2}%
  \BibitemOpen
  \bibfield  {author} {\bibinfo {author} {\bibnamefont {{LIGO Scientific
  Collaboration and Virgo Collaboration}}},\ }\href
  {https://git.ligo.org/lscsoft/lalsuite/tree/lalinference_o2} {\enquote
  {\bibinfo {title} {{LALSuite},
  https://git.ligo.org/lscsoft/lalsuite/tree/lalinference\_o2},}\ } (\bibinfo
  {year} {2017})\BibitemShut {NoStop}%
\bibitem [{\citenamefont {Littenberg}\ and\ \citenamefont
  {Cornish}(2015)}]{Littenberg:2014oda}%
  \BibitemOpen
  \bibfield  {author} {\bibinfo {author} {\bibfnamefont {T.~B.}\ \bibnamefont
  {Littenberg}}\ and\ \bibinfo {author} {\bibfnamefont {N.~J.}\ \bibnamefont
  {Cornish}},\ }\href {\doibase 10.1103/PhysRevD.91.084034} {\bibfield
  {journal} {\bibinfo  {journal} {\prd}\ }\textbf {\bibinfo {volume} {91}},\
  \bibinfo {pages} {084034} (\bibinfo {year} {2015})},\ \Eprint
  {http://arxiv.org/abs/1410.3852} {arXiv:1410.3852 [gr-qc]} \BibitemShut
  {NoStop}%
\bibitem [{\citenamefont {Cornish}\ and\ \citenamefont
  {Littenberg}(2015)}]{CoLi2015}%
  \BibitemOpen
  \bibfield  {author} {\bibinfo {author} {\bibfnamefont {N.~J.}\ \bibnamefont
  {Cornish}}\ and\ \bibinfo {author} {\bibfnamefont {T.~B.}\ \bibnamefont
  {Littenberg}},\ }\href@noop {} {\bibfield  {journal} {\bibinfo  {journal}
  {Class. Quant. Grav.}\ }\textbf {\bibinfo {volume} {32}},\ \bibinfo {pages}
  {135012} (\bibinfo {year} {2015})}\BibitemShut {NoStop}%
\bibitem [{\citenamefont {Farr}\ \emph {et~al.}(2015)\citenamefont {Farr},
  \citenamefont {Farr},\ and\ \citenamefont
  {Littenberg}}]{SplineCalMarg-T1400682}%
  \BibitemOpen
  \bibfield  {author} {\bibinfo {author} {\bibfnamefont {W.~M.}\ \bibnamefont
  {Farr}}, \bibinfo {author} {\bibfnamefont {B.}~\bibnamefont {Farr}}, \ and\
  \bibinfo {author} {\bibfnamefont {T.}~\bibnamefont {Littenberg}},\ }\href
  {https://dcc.ligo.org/LIGO-T1400682/public} {\emph {\bibinfo {title}
  {Modelling Calibration Errors In CBC Waveforms}}},\ \bibinfo {type} {Tech.
  Rep.}\ \bibinfo {number} {{LIGO}-T1400682}\ (\bibinfo  {institution} {{LIGO}
  Project},\ \bibinfo {year} {2015})\BibitemShut {NoStop}%
\bibitem [{\citenamefont {Tauris}\ \emph {et~al.}(2017)\citenamefont {Tauris}
  \emph {et~al.}}]{Tauris:2017omb}%
  \BibitemOpen
  \bibfield  {author} {\bibinfo {author} {\bibfnamefont {T.~M.}\ \bibnamefont
  {Tauris}} \emph {et~al.},\ }\href {\doibase 10.3847/1538-4357/aa7e89}
  {\bibfield  {journal} {\bibinfo  {journal} {Astrophys. J.}\ }\textbf
  {\bibinfo {volume} {846}},\ \bibinfo {pages} {170} (\bibinfo {year}
  {2017})},\ \Eprint {http://arxiv.org/abs/1706.09438} {arXiv:1706.09438
  [astro-ph.HE]} \BibitemShut {NoStop}%
\bibitem [{\citenamefont {Schmidt}\ \emph {et~al.}(2012)\citenamefont
  {Schmidt}, \citenamefont {Hannam},\ and\ \citenamefont
  {Husa}}]{Schmidt:2012rh}%
  \BibitemOpen
  \bibfield  {author} {\bibinfo {author} {\bibfnamefont {P.}~\bibnamefont
  {Schmidt}}, \bibinfo {author} {\bibfnamefont {M.}~\bibnamefont {Hannam}}, \
  and\ \bibinfo {author} {\bibfnamefont {S.}~\bibnamefont {Husa}},\ }\href
  {\doibase 10.1103/PhysRevD.86.104063} {\bibfield  {journal} {\bibinfo
  {journal} {\prd}\ }\textbf {\bibinfo {volume} {86}},\ \bibinfo {pages}
  {104063} (\bibinfo {year} {2012})},\ \Eprint {http://arxiv.org/abs/1207.3088}
  {arXiv:1207.3088 [gr-qc]} \BibitemShut {NoStop}%
\bibitem [{\citenamefont {Hannam}\ \emph {et~al.}(2014)\citenamefont {Hannam},
  \citenamefont {Schmidt}, \citenamefont {Boh{\'e}}, \citenamefont {Haegel},
  \citenamefont {Husa}, \citenamefont {Ohme}, \citenamefont {Pratten},\ and\
  \citenamefont {P{\"u}rrer}}]{Hannam:2013oca}%
  \BibitemOpen
  \bibfield  {author} {\bibinfo {author} {\bibfnamefont {M.}~\bibnamefont
  {Hannam}}, \bibinfo {author} {\bibfnamefont {P.}~\bibnamefont {Schmidt}},
  \bibinfo {author} {\bibfnamefont {A.}~\bibnamefont {Boh{\'e}}}, \bibinfo
  {author} {\bibfnamefont {L.}~\bibnamefont {Haegel}}, \bibinfo {author}
  {\bibfnamefont {S.}~\bibnamefont {Husa}}, \bibinfo {author} {\bibfnamefont
  {F.}~\bibnamefont {Ohme}}, \bibinfo {author} {\bibfnamefont {G.}~\bibnamefont
  {Pratten}}, \ and\ \bibinfo {author} {\bibfnamefont {M.}~\bibnamefont
  {P{\"u}rrer}},\ }\href {\doibase 10.1103/PhysRevLett.113.151101} {\bibfield
  {journal} {\bibinfo  {journal} {Phys. Rev. Lett.}\ }\textbf {\bibinfo
  {volume} {113}},\ \bibinfo {pages} {151101} (\bibinfo {year} {2014})},\
  \Eprint {http://arxiv.org/abs/1308.3271} {arXiv:1308.3271 [gr-qc]}
  \BibitemShut {NoStop}%
\bibitem [{\citenamefont {Schmidt}\ \emph {et~al.}(2015)\citenamefont
  {Schmidt}, \citenamefont {Ohme},\ and\ \citenamefont
  {Hannam}}]{Schmidt:2014iyl}%
  \BibitemOpen
  \bibfield  {author} {\bibinfo {author} {\bibfnamefont {P.}~\bibnamefont
  {Schmidt}}, \bibinfo {author} {\bibfnamefont {F.}~\bibnamefont {Ohme}}, \
  and\ \bibinfo {author} {\bibfnamefont {M.}~\bibnamefont {Hannam}},\ }\href
  {\doibase 10.1103/PhysRevD.91.024043} {\bibfield  {journal} {\bibinfo
  {journal} {Phys. Rev. D}\ }\textbf {\bibinfo {volume} {91}},\ \bibinfo
  {pages} {024043} (\bibinfo {year} {2015})},\ \Eprint
  {http://arxiv.org/abs/1408.1810} {arXiv:1408.1810 [gr-qc]} \BibitemShut
  {NoStop}%
\bibitem [{\citenamefont {Husa}\ \emph {et~al.}(2016)\citenamefont {Husa},
  \citenamefont {Khan}, \citenamefont {Hannam}, \citenamefont {P{\"u}rrer},
  \citenamefont {Ohme}, \citenamefont {Forteza},\ and\ \citenamefont
  {Boh{\'e}}}]{Husa:2015iqa}%
  \BibitemOpen
  \bibfield  {author} {\bibinfo {author} {\bibfnamefont {S.}~\bibnamefont
  {Husa}}, \bibinfo {author} {\bibfnamefont {S.}~\bibnamefont {Khan}}, \bibinfo
  {author} {\bibfnamefont {M.}~\bibnamefont {Hannam}}, \bibinfo {author}
  {\bibfnamefont {M.}~\bibnamefont {P{\"u}rrer}}, \bibinfo {author}
  {\bibfnamefont {F.}~\bibnamefont {Ohme}}, \bibinfo {author} {\bibfnamefont
  {X.~J.}\ \bibnamefont {Forteza}}, \ and\ \bibinfo {author} {\bibfnamefont
  {A.}~\bibnamefont {Boh{\'e}}},\ }\href {\doibase 10.1103/PhysRevD.93.044006}
  {\bibfield  {journal} {\bibinfo  {journal} {Phys. Rev. D}\ }\textbf {\bibinfo
  {volume} {93}},\ \bibinfo {pages} {044006} (\bibinfo {year} {2016})},\
  \Eprint {http://arxiv.org/abs/1508.07250} {arXiv:1508.07250 [gr-qc]}
  \BibitemShut {NoStop}%
\bibitem [{\citenamefont {Khan}\ \emph {et~al.}(2016)\citenamefont {Khan},
  \citenamefont {Husa}, \citenamefont {Hannam}, \citenamefont {Ohme},
  \citenamefont {P{\"{u}}rrer}, \citenamefont {Forteza},\ and\ \citenamefont
  {Boh{\'e}}}]{Khan:2015jqa}%
  \BibitemOpen
  \bibfield  {author} {\bibinfo {author} {\bibfnamefont {S.}~\bibnamefont
  {Khan}}, \bibinfo {author} {\bibfnamefont {S.}~\bibnamefont {Husa}}, \bibinfo
  {author} {\bibfnamefont {M.}~\bibnamefont {Hannam}}, \bibinfo {author}
  {\bibfnamefont {F.}~\bibnamefont {Ohme}}, \bibinfo {author} {\bibfnamefont
  {M.}~\bibnamefont {P{\"{u}}rrer}}, \bibinfo {author} {\bibfnamefont {X.~J.}\
  \bibnamefont {Forteza}}, \ and\ \bibinfo {author} {\bibfnamefont
  {A.}~\bibnamefont {Boh{\'e}}},\ }\href {\doibase 10.1103/PhysRevD.93.044007}
  {\bibfield  {journal} {\bibinfo  {journal} {Phys. Rev. D}\ }\textbf {\bibinfo
  {volume} {93}},\ \bibinfo {pages} {044007} (\bibinfo {year} {2016})},\
  \Eprint {http://arxiv.org/abs/1508.07253} {arXiv:1508.07253 [gr-qc]}
  \BibitemShut {NoStop}%
\bibitem [{\citenamefont {Dietrich}\ \emph {et~al.}(2017)\citenamefont
  {Dietrich}, \citenamefont {Bernuzzi},\ and\ \citenamefont
  {Tichy}}]{Dietrich:2017aum}%
  \BibitemOpen
  \bibfield  {author} {\bibinfo {author} {\bibfnamefont {T.}~\bibnamefont
  {Dietrich}}, \bibinfo {author} {\bibfnamefont {S.}~\bibnamefont {Bernuzzi}},
  \ and\ \bibinfo {author} {\bibfnamefont {W.}~\bibnamefont {Tichy}},\ }\href
  {\doibase 10.1103/PhysRevD.96.121501} {\bibfield  {journal} {\bibinfo
  {journal} {\prd}\ }\textbf {\bibinfo {volume} {96}},\ \bibinfo {pages}
  {121501} (\bibinfo {year} {2017})},\ \Eprint
  {http://arxiv.org/abs/1706.02969} {arXiv:1706.02969 [gr-qc]} \BibitemShut
  {NoStop}%
\bibitem [{\citenamefont {Boh{\'e}}\ \emph {et~al.}(2017)\citenamefont
  {Boh{\'e}} \emph {et~al.}}]{Bohe:2016gbl}%
  \BibitemOpen
  \bibfield  {author} {\bibinfo {author} {\bibfnamefont {A.}~\bibnamefont
  {Boh{\'e}}} \emph {et~al.},\ }\href {\doibase 10.1103/PhysRevD.95.044028}
  {\bibfield  {journal} {\bibinfo  {journal} {\prd}\ }\textbf {\bibinfo
  {volume} {95}},\ \bibinfo {pages} {044028} (\bibinfo {year} {2017})},\
  \Eprint {http://arxiv.org/abs/1611.03703} {arXiv:1611.03703 [gr-qc]}
  \BibitemShut {NoStop}%
\bibitem [{\citenamefont {Bernuzzi}\ \emph {et~al.}(2015)\citenamefont
  {Bernuzzi}, \citenamefont {Nagar}, \citenamefont {Dietrich},\ and\
  \citenamefont {Damour}}]{Bernuzzi:2014owa}%
  \BibitemOpen
  \bibfield  {author} {\bibinfo {author} {\bibfnamefont {S.}~\bibnamefont
  {Bernuzzi}}, \bibinfo {author} {\bibfnamefont {A.}~\bibnamefont {Nagar}},
  \bibinfo {author} {\bibfnamefont {T.}~\bibnamefont {Dietrich}}, \ and\
  \bibinfo {author} {\bibfnamefont {T.}~\bibnamefont {Damour}},\ }\href
  {\doibase 10.1103/PhysRevLett.114.161103} {\bibfield  {journal} {\bibinfo
  {journal} {Phys. Rev. Lett.}\ }\textbf {\bibinfo {volume} {114}},\ \bibinfo
  {pages} {161103} (\bibinfo {year} {2015})},\ \Eprint
  {http://arxiv.org/abs/1412.4553} {arXiv:1412.4553 [gr-qc]} \BibitemShut
  {NoStop}%
\bibitem [{\citenamefont {Lackey}\ \emph {et~al.}(2017)\citenamefont {Lackey},
  \citenamefont {Bernuzzi}, \citenamefont {Galley}, \citenamefont {Meidam},\
  and\ \citenamefont {Van Den~Broeck}}]{Lackey:2016krb}%
  \BibitemOpen
  \bibfield  {author} {\bibinfo {author} {\bibfnamefont {B.~D.}\ \bibnamefont
  {Lackey}}, \bibinfo {author} {\bibfnamefont {S.}~\bibnamefont {Bernuzzi}},
  \bibinfo {author} {\bibfnamefont {C.~R.}\ \bibnamefont {Galley}}, \bibinfo
  {author} {\bibfnamefont {J.}~\bibnamefont {Meidam}}, \ and\ \bibinfo {author}
  {\bibfnamefont {C.}~\bibnamefont {Van Den~Broeck}},\ }\href {\doibase
  10.1103/PhysRevD.95.104036} {\bibfield  {journal} {\bibinfo  {journal}
  {\prd}\ }\textbf {\bibinfo {volume} {95}},\ \bibinfo {pages} {104036}
  (\bibinfo {year} {2017})},\ \Eprint {http://arxiv.org/abs/1610.04742}
  {arXiv:1610.04742 [gr-qc]} \BibitemShut {NoStop}%
\bibitem [{\citenamefont {Hinderer}\ \emph {et~al.}(2016)\citenamefont
  {Hinderer} \emph {et~al.}}]{Hinderer:2016eia}%
  \BibitemOpen
  \bibfield  {author} {\bibinfo {author} {\bibfnamefont {T.}~\bibnamefont
  {Hinderer}} \emph {et~al.},\ }\href {\doibase 10.1103/PhysRevLett.116.181101}
  {\bibfield  {journal} {\bibinfo  {journal} {Phys. Rev. Lett.}\ }\textbf
  {\bibinfo {volume} {116}},\ \bibinfo {pages} {181101} (\bibinfo {year}
  {2016})},\ \Eprint {http://arxiv.org/abs/1602.00599} {arXiv:1602.00599
  [gr-qc]} \BibitemShut {NoStop}%
\bibitem [{\citenamefont {Steinhoff}\ \emph {et~al.}(2016)\citenamefont
  {Steinhoff}, \citenamefont {Hinderer}, \citenamefont {Buonanno},\ and\
  \citenamefont {Taracchini}}]{Steinhoff:2016rfi}%
  \BibitemOpen
  \bibfield  {author} {\bibinfo {author} {\bibfnamefont {J.}~\bibnamefont
  {Steinhoff}}, \bibinfo {author} {\bibfnamefont {T.}~\bibnamefont {Hinderer}},
  \bibinfo {author} {\bibfnamefont {A.}~\bibnamefont {Buonanno}}, \ and\
  \bibinfo {author} {\bibfnamefont {A.}~\bibnamefont {Taracchini}},\ }\href
  {\doibase 10.1103/PhysRevD.94.104028} {\bibfield  {journal} {\bibinfo
  {journal} {\prd}\ }\textbf {\bibinfo {volume} {94}},\ \bibinfo {pages}
  {104028} (\bibinfo {year} {2016})},\ \Eprint
  {http://arxiv.org/abs/1608.01907} {arXiv:1608.01907 [gr-qc]} \BibitemShut
  {NoStop}%
\bibitem [{\citenamefont {Sathyaprakash}\ and\ \citenamefont
  {Dhurandhar}(1991)}]{Sathyaprakash:1991mt}%
  \BibitemOpen
  \bibfield  {author} {\bibinfo {author} {\bibfnamefont {B.~S.}\ \bibnamefont
  {Sathyaprakash}}\ and\ \bibinfo {author} {\bibfnamefont {S.~V.}\ \bibnamefont
  {Dhurandhar}},\ }\href {\doibase 10.1103/PhysRevD.44.3819} {\bibfield
  {journal} {\bibinfo  {journal} {Phys. Rev. D}\ }\textbf {\bibinfo {volume}
  {44}},\ \bibinfo {pages} {3819} (\bibinfo {year} {1991})}\BibitemShut
  {NoStop}%
\bibitem [{\citenamefont {Boh{\'e}}\ \emph {et~al.}(2013)\citenamefont
  {Boh{\'e}}, \citenamefont {Marsat},\ and\ \citenamefont
  {Blanchet}}]{Bohe:2013cla}%
  \BibitemOpen
  \bibfield  {author} {\bibinfo {author} {\bibfnamefont {A.}~\bibnamefont
  {Boh{\'e}}}, \bibinfo {author} {\bibfnamefont {S.}~\bibnamefont {Marsat}}, \
  and\ \bibinfo {author} {\bibfnamefont {L.}~\bibnamefont {Blanchet}},\ }\href
  {\doibase 10.1088/0264-9381/30/13/135009} {\bibfield  {journal} {\bibinfo
  {journal} {Class. Quant. Grav.}\ }\textbf {\bibinfo {volume} {30}},\ \bibinfo
  {pages} {135009} (\bibinfo {year} {2013})},\ \Eprint
  {http://arxiv.org/abs/1303.7412} {arXiv:1303.7412 [gr-qc]} \BibitemShut
  {NoStop}%
\bibitem [{\citenamefont {Arun}\ \emph {et~al.}(2009)\citenamefont {Arun},
  \citenamefont {Buonanno}, \citenamefont {Faye},\ and\ \citenamefont
  {Ochsner}}]{Arun:2008kb}%
  \BibitemOpen
  \bibfield  {author} {\bibinfo {author} {\bibfnamefont {K.}~\bibnamefont
  {Arun}}, \bibinfo {author} {\bibfnamefont {A.}~\bibnamefont {Buonanno}},
  \bibinfo {author} {\bibfnamefont {G.}~\bibnamefont {Faye}}, \ and\ \bibinfo
  {author} {\bibfnamefont {E.}~\bibnamefont {Ochsner}},\ }\href@noop {}
  {\bibfield  {journal} {\bibinfo  {journal} {\prd}\ }\textbf {\bibinfo
  {volume} {79}},\ \bibinfo {pages} {104023} (\bibinfo {year} {2009})},\
  \Eprint {http://arxiv.org/abs/0810.5336} {arXiv:0810.5336 [gr-qc]}
  \BibitemShut {NoStop}%
\bibitem [{\citenamefont {Mikoczi}\ \emph {et~al.}(2005)\citenamefont
  {Mikoczi}, \citenamefont {Vasuth},\ and\ \citenamefont
  {Gergely}}]{Mikoczi:2005dn}%
  \BibitemOpen
  \bibfield  {author} {\bibinfo {author} {\bibfnamefont {B.}~\bibnamefont
  {Mikoczi}}, \bibinfo {author} {\bibfnamefont {M.}~\bibnamefont {Vasuth}}, \
  and\ \bibinfo {author} {\bibfnamefont {L.~A.}\ \bibnamefont {Gergely}},\
  }\href {\doibase 10.1103/PhysRevD.71.124043} {\bibfield  {journal} {\bibinfo
  {journal} {\prd}\ }\textbf {\bibinfo {volume} {71}},\ \bibinfo {pages}
  {124043} (\bibinfo {year} {2005})},\ \Eprint
  {http://arxiv.org/abs/astro-ph/0504538} {arXiv:astro-ph/0504538 [astro-ph]}
  \BibitemShut {NoStop}%
\bibitem [{\citenamefont {Mishra}\ \emph {et~al.}(2016)\citenamefont {Mishra},
  \citenamefont {Kela}, \citenamefont {Arun},\ and\ \citenamefont
  {Faye}}]{Mishra:2016whh}%
  \BibitemOpen
  \bibfield  {author} {\bibinfo {author} {\bibfnamefont {C.~K.}\ \bibnamefont
  {Mishra}}, \bibinfo {author} {\bibfnamefont {A.}~\bibnamefont {Kela}},
  \bibinfo {author} {\bibfnamefont {K.~G.}\ \bibnamefont {Arun}}, \ and\
  \bibinfo {author} {\bibfnamefont {G.}~\bibnamefont {Faye}},\ }\href {\doibase
  10.1103/PhysRevD.93.084054} {\bibfield  {journal} {\bibinfo  {journal}
  {\prd}\ }\textbf {\bibinfo {volume} {93}},\ \bibinfo {pages} {084054}
  (\bibinfo {year} {2016})},\ \Eprint {http://arxiv.org/abs/1601.05588}
  {arXiv:1601.05588 [gr-qc]} \BibitemShut {NoStop}%
\bibitem [{\citenamefont {{Taracchini}}\ \emph {et~al.}(2014)\citenamefont
  {{Taracchini}}, \citenamefont {{Buonanno}}, \citenamefont {{Pan}},
  \citenamefont {{Hinderer}}, \citenamefont {{Boyle}}, \citenamefont
  {{Hemberger}}, \citenamefont {{Kidder}}, \citenamefont {{Lovelace}},
  \citenamefont {{Mrou{\'e}}}, \citenamefont {{Pfeiffer}}, \citenamefont
  {{Scheel}}, \citenamefont {{Szil{\'a}gyi}}, \citenamefont {{Taylor}},\ and\
  \citenamefont {{Zenginoglu}}}]{Taracchini:2013}%
  \BibitemOpen
  \bibfield  {author} {\bibinfo {author} {\bibfnamefont {A.}~\bibnamefont
  {{Taracchini}}}, \bibinfo {author} {\bibfnamefont {A.}~\bibnamefont
  {{Buonanno}}}, \bibinfo {author} {\bibfnamefont {Y.}~\bibnamefont {{Pan}}},
  \bibinfo {author} {\bibfnamefont {T.}~\bibnamefont {{Hinderer}}}, \bibinfo
  {author} {\bibfnamefont {M.}~\bibnamefont {{Boyle}}}, \bibinfo {author}
  {\bibfnamefont {D.~A.}\ \bibnamefont {{Hemberger}}}, \bibinfo {author}
  {\bibfnamefont {L.~E.}\ \bibnamefont {{Kidder}}}, \bibinfo {author}
  {\bibfnamefont {G.}~\bibnamefont {{Lovelace}}}, \bibinfo {author}
  {\bibfnamefont {A.~H.}\ \bibnamefont {{Mrou{\'e}}}}, \bibinfo {author}
  {\bibfnamefont {H.~P.}\ \bibnamefont {{Pfeiffer}}}, \bibinfo {author}
  {\bibfnamefont {M.~A.}\ \bibnamefont {{Scheel}}}, \bibinfo {author}
  {\bibfnamefont {B.}~\bibnamefont {{Szil{\'a}gyi}}}, \bibinfo {author}
  {\bibfnamefont {N.~W.}\ \bibnamefont {{Taylor}}}, \ and\ \bibinfo {author}
  {\bibfnamefont {A.}~\bibnamefont {{Zenginoglu}}},\ }\href {\doibase
  10.1103/PhysRevD.89.061502} {\bibfield  {journal} {\bibinfo  {journal}
  {\prd}\ }\textbf {\bibinfo {volume} {89}},\ \bibinfo {eid} {061502} (\bibinfo
  {year} {2014})},\ \Eprint {http://arxiv.org/abs/1311.2544} {arXiv:1311.2544
  [gr-qc]} \BibitemShut {NoStop}%
\bibitem [{\citenamefont {Barausse}\ and\ \citenamefont
  {Buonanno}(2010)}]{Barausse:2009xi}%
  \BibitemOpen
  \bibfield  {author} {\bibinfo {author} {\bibfnamefont {E.}~\bibnamefont
  {Barausse}}\ and\ \bibinfo {author} {\bibfnamefont {A.}~\bibnamefont
  {Buonanno}},\ }\href {\doibase 10.1103/PhysRevD.81.084024} {\bibfield
  {journal} {\bibinfo  {journal} {Phys. Rev. D}\ }\textbf {\bibinfo {volume}
  {81}},\ \bibinfo {pages} {084024} (\bibinfo {year} {2010})},\ \Eprint
  {http://arxiv.org/abs/0912.3517} {arXiv:0912.3517 [gr-qc]} \BibitemShut
  {NoStop}%
\bibitem [{\citenamefont {Buonanno}\ and\ \citenamefont
  {Damour}(2000)}]{Buonanno:2000ef}%
  \BibitemOpen
  \bibfield  {author} {\bibinfo {author} {\bibfnamefont {A.}~\bibnamefont
  {Buonanno}}\ and\ \bibinfo {author} {\bibfnamefont {T.}~\bibnamefont
  {Damour}},\ }\href {\doibase 10.1103/PhysRevD.62.064015} {\bibfield
  {journal} {\bibinfo  {journal} {Phys. Rev. D}\ }\textbf {\bibinfo {volume}
  {62}},\ \bibinfo {pages} {064015} (\bibinfo {year} {2000})},\ \Eprint
  {http://arxiv.org/abs/gr-qc/0001013} {arXiv:gr-qc/0001013 [gr-qc]}
  \BibitemShut {NoStop}%
\bibitem [{\citenamefont {Buonanno}\ and\ \citenamefont
  {Damour}(1999)}]{BuonannoDamour:1999}%
  \BibitemOpen
  \bibfield  {author} {\bibinfo {author} {\bibfnamefont {A.}~\bibnamefont
  {Buonanno}}\ and\ \bibinfo {author} {\bibfnamefont {T.}~\bibnamefont
  {Damour}},\ }\href {\doibase 10.1103/PhysRevD.59.084006} {\bibfield
  {journal} {\bibinfo  {journal} {Phys. Rev. D}\ }\textbf {\bibinfo {volume}
  {59}},\ \bibinfo {pages} {084006} (\bibinfo {year} {1999})},\ \Eprint
  {http://arxiv.org/abs/gr-qc/9811091} {arXiv:gr-qc/9811091 [gr-qc]}
  \BibitemShut {NoStop}%
\bibitem [{\citenamefont {Wade}\ \emph {et~al.}(2014)\citenamefont {Wade},
  \citenamefont {Creighton}, \citenamefont {Ochsner}, \citenamefont {Lackey},
  \citenamefont {Farr}, \citenamefont {Littenberg},\ and\ \citenamefont
  {Raymond}}]{Wade:2014vqa}%
  \BibitemOpen
  \bibfield  {author} {\bibinfo {author} {\bibfnamefont {L.}~\bibnamefont
  {Wade}}, \bibinfo {author} {\bibfnamefont {J.~D.~E.}\ \bibnamefont
  {Creighton}}, \bibinfo {author} {\bibfnamefont {E.}~\bibnamefont {Ochsner}},
  \bibinfo {author} {\bibfnamefont {B.~D.}\ \bibnamefont {Lackey}}, \bibinfo
  {author} {\bibfnamefont {B.~F.}\ \bibnamefont {Farr}}, \bibinfo {author}
  {\bibfnamefont {T.~B.}\ \bibnamefont {Littenberg}}, \ and\ \bibinfo {author}
  {\bibfnamefont {V.}~\bibnamefont {Raymond}},\ }\href {\doibase
  10.1103/PhysRevD.89.103012} {\bibfield  {journal} {\bibinfo  {journal}
  {\prd}\ }\textbf {\bibinfo {volume} {89}},\ \bibinfo {pages} {103012}
  (\bibinfo {year} {2014})},\ \Eprint {http://arxiv.org/abs/1402.5156}
  {arXiv:1402.5156 [gr-qc]} \BibitemShut {NoStop}%
\bibitem [{\citenamefont {Dietrich}\ \emph {et~al.}(2018)\citenamefont
  {Dietrich} \emph {et~al.}}]{Dietrich:2018uni}%
  \BibitemOpen
  \bibfield  {author} {\bibinfo {author} {\bibfnamefont {T.}~\bibnamefont
  {Dietrich}} \emph {et~al.},\ }\href@noop {} {\  (\bibinfo {year} {2018})},\
  \Eprint {http://arxiv.org/abs/1804.02235} {arXiv:1804.02235 [gr-qc]}
  \BibitemShut {NoStop}%
\bibitem [{\citenamefont {Yagi}\ and\ \citenamefont
  {Yunes}(2013{\natexlab{a}})}]{Yagi:2013bca}%
  \BibitemOpen
  \bibfield  {author} {\bibinfo {author} {\bibfnamefont {K.}~\bibnamefont
  {Yagi}}\ and\ \bibinfo {author} {\bibfnamefont {N.}~\bibnamefont {Yunes}},\
  }\href {\doibase 10.1126/science.1236462} {\bibfield  {journal} {\bibinfo
  {journal} {Science}\ }\textbf {\bibinfo {volume} {341}},\ \bibinfo {pages}
  {365} (\bibinfo {year} {2013}{\natexlab{a}})},\ \Eprint
  {http://arxiv.org/abs/1302.4499} {arXiv:1302.4499 [gr-qc]} \BibitemShut
  {NoStop}%
\bibitem [{\citenamefont {Yagi}\ and\ \citenamefont
  {Yunes}(2013{\natexlab{b}})}]{Yagi:2013awa}%
  \BibitemOpen
  \bibfield  {author} {\bibinfo {author} {\bibfnamefont {K.}~\bibnamefont
  {Yagi}}\ and\ \bibinfo {author} {\bibfnamefont {N.}~\bibnamefont {Yunes}},\
  }\href {\doibase 10.1103/PhysRevD.88.023009} {\bibfield  {journal} {\bibinfo
  {journal} {\prd}\ }\textbf {\bibinfo {volume} {88}},\ \bibinfo {pages}
  {023009} (\bibinfo {year} {2013}{\natexlab{b}})},\ \Eprint
  {http://arxiv.org/abs/1303.1528} {arXiv:1303.1528 [gr-qc]} \BibitemShut
  {NoStop}%
\bibitem [{\citenamefont {Chatziioannou}\ \emph {et~al.}(2015)\citenamefont
  {Chatziioannou}, \citenamefont {Yagi}, \citenamefont {Klein}, \citenamefont
  {Cornish},\ and\ \citenamefont {Yunes}}]{Chatziioannou:2015uea}%
  \BibitemOpen
  \bibfield  {author} {\bibinfo {author} {\bibfnamefont {K.}~\bibnamefont
  {Chatziioannou}}, \bibinfo {author} {\bibfnamefont {K.}~\bibnamefont {Yagi}},
  \bibinfo {author} {\bibfnamefont {A.}~\bibnamefont {Klein}}, \bibinfo
  {author} {\bibfnamefont {N.}~\bibnamefont {Cornish}}, \ and\ \bibinfo
  {author} {\bibfnamefont {N.}~\bibnamefont {Yunes}},\ }\href {\doibase
  10.1103/PhysRevD.92.104008} {\bibfield  {journal} {\bibinfo  {journal}
  {\prd}\ }\textbf {\bibinfo {volume} {92}},\ \bibinfo {pages} {104008}
  (\bibinfo {year} {2015})},\ \Eprint {http://arxiv.org/abs/1508.02062}
  {arXiv:1508.02062 [gr-qc]} \BibitemShut {NoStop}%
\bibitem [{\citenamefont {Yagi}\ and\ \citenamefont
  {Yunes}(2017)}]{Yagi:2016bkt}%
  \BibitemOpen
  \bibfield  {author} {\bibinfo {author} {\bibfnamefont {K.}~\bibnamefont
  {Yagi}}\ and\ \bibinfo {author} {\bibfnamefont {N.}~\bibnamefont {Yunes}},\
  }\href {\doibase 10.1016/j.physrep.2017.03.002} {\bibfield  {journal}
  {\bibinfo  {journal} {Phys. Rept.}\ }\textbf {\bibinfo {volume} {681}},\
  \bibinfo {pages} {1} (\bibinfo {year} {2017})},\ \Eprint
  {http://arxiv.org/abs/1608.02582} {arXiv:1608.02582 [gr-qc]} \BibitemShut
  {NoStop}%
\bibitem [{\citenamefont {Yagi}\ and\ \citenamefont
  {Yunes}(2016)}]{Yagi:2015pkc}%
  \BibitemOpen
  \bibfield  {author} {\bibinfo {author} {\bibfnamefont {K.}~\bibnamefont
  {Yagi}}\ and\ \bibinfo {author} {\bibfnamefont {N.}~\bibnamefont {Yunes}},\
  }\href {\doibase 10.1088/0264-9381/33/13/13LT01} {\bibfield  {journal}
  {\bibinfo  {journal} {Class. Quant. Grav.}\ }\textbf {\bibinfo {volume}
  {33}},\ \bibinfo {pages} {13LT01} (\bibinfo {year} {2016})},\ \Eprint
  {http://arxiv.org/abs/1512.02639} {arXiv:1512.02639 [gr-qc]} \BibitemShut
  {NoStop}%
\bibitem [{\citenamefont {Chatziioannou}\ \emph {et~al.}(2018)\citenamefont
  {Chatziioannou}, \citenamefont {Haster},\ and\ \citenamefont
  {Zimmerman}}]{Chatziioannou:2018vzf}%
  \BibitemOpen
  \bibfield  {author} {\bibinfo {author} {\bibfnamefont {K.}~\bibnamefont
  {Chatziioannou}}, \bibinfo {author} {\bibfnamefont {C.-J.}\ \bibnamefont
  {Haster}}, \ and\ \bibinfo {author} {\bibfnamefont {A.}~\bibnamefont
  {Zimmerman}},\ }\href {\doibase 10.1103/PhysRevD.97.104036} {\bibfield
  {journal} {\bibinfo  {journal} {Phys. Rev. D}\ }\textbf {\bibinfo {volume}
  {97}},\ \bibinfo {pages} {104036} (\bibinfo {year} {2018})},\ \Eprint
  {http://arxiv.org/abs/1804.03221} {arXiv:1804.03221 [gr-qc]} \BibitemShut
  {NoStop}%
\bibitem [{\citenamefont {Maselli}\ \emph {et~al.}(2013)\citenamefont
  {Maselli}, \citenamefont {Cardoso}, \citenamefont {Ferrari}, \citenamefont
  {Gualtieri},\ and\ \citenamefont {Pani}}]{Maselli:2013mva}%
  \BibitemOpen
  \bibfield  {author} {\bibinfo {author} {\bibfnamefont {A.}~\bibnamefont
  {Maselli}}, \bibinfo {author} {\bibfnamefont {V.}~\bibnamefont {Cardoso}},
  \bibinfo {author} {\bibfnamefont {V.}~\bibnamefont {Ferrari}}, \bibinfo
  {author} {\bibfnamefont {L.}~\bibnamefont {Gualtieri}}, \ and\ \bibinfo
  {author} {\bibfnamefont {P.}~\bibnamefont {Pani}},\ }\href {\doibase
  10.1103/PhysRevD.88.023007} {\bibfield  {journal} {\bibinfo  {journal}
  {\prd}\ }\textbf {\bibinfo {volume} {88}},\ \bibinfo {pages} {023007}
  (\bibinfo {year} {2013})},\ \Eprint {http://arxiv.org/abs/1304.2052}
  {arXiv:1304.2052 [gr-qc]} \BibitemShut {NoStop}%
\bibitem [{\citenamefont {Urbanec}\ \emph {et~al.}(2013)\citenamefont
  {Urbanec}, \citenamefont {Miller},\ and\ \citenamefont
  {Stuchlik}}]{Urbanec:2013fs}%
  \BibitemOpen
  \bibfield  {author} {\bibinfo {author} {\bibfnamefont {M.}~\bibnamefont
  {Urbanec}}, \bibinfo {author} {\bibfnamefont {J.~C.}\ \bibnamefont {Miller}},
  \ and\ \bibinfo {author} {\bibfnamefont {Z.}~\bibnamefont {Stuchlik}},\
  }\href {\doibase 10.1093/mnras/stt858} {\bibfield  {journal} {\bibinfo
  {journal} {Mon. Not. Roy. Astron. Soc.}\ }\textbf {\bibinfo {volume} {433}},\
  \bibinfo {pages} {1903} (\bibinfo {year} {2013})},\ \Eprint
  {http://arxiv.org/abs/1301.5925} {arXiv:1301.5925 [astro-ph.SR]} \BibitemShut
  {NoStop}%
\bibitem [{\citenamefont {Lindblom}(2018)}]{Lindblom:2018rfr}%
  \BibitemOpen
  \bibfield  {author} {\bibinfo {author} {\bibfnamefont {L.}~\bibnamefont
  {Lindblom}},\ }\href@noop {} {\  (\bibinfo {year} {2018})},\ \Eprint
  {http://arxiv.org/abs/1804.04072} {arXiv:1804.04072 [astro-ph.HE]}
  \BibitemShut {NoStop}%
\bibitem [{\citenamefont {Lindblom}(2010)}]{PhysRevD.82.103011}%
  \BibitemOpen
  \bibfield  {author} {\bibinfo {author} {\bibfnamefont {L.}~\bibnamefont
  {Lindblom}},\ }\href {\doibase 10.1103/PhysRevD.82.103011} {\bibfield
  {journal} {\bibinfo  {journal} {Phys. Rev. D}\ }\textbf {\bibinfo {volume}
  {82}},\ \bibinfo {pages} {103011} (\bibinfo {year} {2010})}\BibitemShut
  {NoStop}%
\bibitem [{\citenamefont {{Lindblom}}\ and\ \citenamefont
  {{Indik}}(2012)}]{2012PhRvD..86h4003L}%
  \BibitemOpen
  \bibfield  {author} {\bibinfo {author} {\bibfnamefont {L.}~\bibnamefont
  {{Lindblom}}}\ and\ \bibinfo {author} {\bibfnamefont {N.~M.}\ \bibnamefont
  {{Indik}}},\ }\href {\doibase 10.1103/PhysRevD.86.084003} {\bibfield
  {journal} {\bibinfo  {journal} {\prd}\ }\textbf {\bibinfo {volume} {86}},\
  \bibinfo {eid} {084003} (\bibinfo {year} {2012})},\ \Eprint
  {http://arxiv.org/abs/1207.3744} {arXiv:1207.3744 [astro-ph.HE]} \BibitemShut
  {NoStop}%
\bibitem [{\citenamefont {Lindblom}\ and\ \citenamefont
  {Indik}(2014)}]{Lindblom:2013kra}%
  \BibitemOpen
  \bibfield  {author} {\bibinfo {author} {\bibfnamefont {L.}~\bibnamefont
  {Lindblom}}\ and\ \bibinfo {author} {\bibfnamefont {N.~M.}\ \bibnamefont
  {Indik}},\ }\href {\doibase 10.1103/PhysRevD.89.064003,
  10.1103/PhysRevD.93.129903} {\bibfield  {journal} {\bibinfo  {journal}
  {\prd}\ }\textbf {\bibinfo {volume} {89}},\ \bibinfo {pages} {064003}
  (\bibinfo {year} {2014})},\ \bibinfo {note} {[Erratum: Phys.
  Rev.D93,no.12,129903(2016)]},\ \Eprint {http://arxiv.org/abs/1310.0803}
  {arXiv:1310.0803 [astro-ph.HE]} \BibitemShut {NoStop}%
\bibitem [{\citenamefont {Douchin}\ and\ \citenamefont
  {Haensel}(2001)}]{Douchin:2001sv}%
  \BibitemOpen
  \bibfield  {author} {\bibinfo {author} {\bibfnamefont {F.}~\bibnamefont
  {Douchin}}\ and\ \bibinfo {author} {\bibfnamefont {P.}~\bibnamefont
  {Haensel}},\ }\href {\doibase 10.1051/0004-6361:20011402} {\bibfield
  {journal} {\bibinfo  {journal} {Astron. Astrophys}\ }\textbf {\bibinfo
  {volume} {380}},\ \bibinfo {pages} {151} (\bibinfo {year} {2001})},\ \Eprint
  {http://arxiv.org/abs/astro-ph/0111092} {arXiv:astro-ph/0111092} \BibitemShut
  {NoStop}%
\bibitem [{\citenamefont {Lattimer}\ and\ \citenamefont
  {Prakash}(2001)}]{Lattimer:2000nx}%
  \BibitemOpen
  \bibfield  {author} {\bibinfo {author} {\bibfnamefont {J.~M.}\ \bibnamefont
  {Lattimer}}\ and\ \bibinfo {author} {\bibfnamefont {M.}~\bibnamefont
  {Prakash}},\ }\href {\doibase 10.1086/319702} {\bibfield  {journal} {\bibinfo
   {journal} {Astrophys. J.}\ }\textbf {\bibinfo {volume} {550}},\ \bibinfo
  {pages} {426} (\bibinfo {year} {2001})},\ \Eprint
  {http://arxiv.org/abs/astro-ph/0002232} {arXiv:astro-ph/0002232 [astro-ph]}
  \BibitemShut {NoStop}%
\bibitem [{\citenamefont {Read}\ \emph {et~al.}(2009)\citenamefont {Read},
  \citenamefont {Lackey}, \citenamefont {Owen},\ and\ \citenamefont
  {Friedman}}]{Read:2008iy}%
  \BibitemOpen
  \bibfield  {author} {\bibinfo {author} {\bibfnamefont {J.~S.}\ \bibnamefont
  {Read}}, \bibinfo {author} {\bibfnamefont {B.~D.}\ \bibnamefont {Lackey}},
  \bibinfo {author} {\bibfnamefont {B.~J.}\ \bibnamefont {Owen}}, \ and\
  \bibinfo {author} {\bibfnamefont {J.~L.}\ \bibnamefont {Friedman}},\ }\href
  {\doibase 10.1103/PhysRevD.79.124032} {\bibfield  {journal} {\bibinfo
  {journal} {Phys. Rev. D}\ }\textbf {\bibinfo {volume} {79}},\ \bibinfo
  {pages} {124032} (\bibinfo {year} {2009})},\ \Eprint
  {http://arxiv.org/abs/0812.2163} {arXiv:0812.2163 [astro-ph]} \BibitemShut
  {NoStop}%
\bibitem [{\citenamefont {Raaijmakers}\ \emph {et~al.}(2018)\citenamefont
  {Raaijmakers}, \citenamefont {Riley},\ and\ \citenamefont
  {Watts}}]{Raaijmakers:2018bln}%
  \BibitemOpen
  \bibfield  {author} {\bibinfo {author} {\bibfnamefont {G.}~\bibnamefont
  {Raaijmakers}}, \bibinfo {author} {\bibfnamefont {T.~E.}\ \bibnamefont
  {Riley}}, \ and\ \bibinfo {author} {\bibfnamefont {A.~L.}\ \bibnamefont
  {Watts}},\ }\href@noop {} {\  (\bibinfo {year} {2018})},\ \Eprint
  {http://arxiv.org/abs/1804.09087} {arXiv:1804.09087 [astro-ph.HE]}
  \BibitemShut {NoStop}%
\bibitem [{\citenamefont {Carney}\ \emph {et~al.}(2018)\citenamefont {Carney},
  \citenamefont {Wade},\ and\ \citenamefont {Irwin}}]{PhysRevD.98.063004}%
  \BibitemOpen
  \bibfield  {author} {\bibinfo {author} {\bibfnamefont {M.~F.}\ \bibnamefont
  {Carney}}, \bibinfo {author} {\bibfnamefont {L.~E.}\ \bibnamefont {Wade}}, \
  and\ \bibinfo {author} {\bibfnamefont {B.~S.}\ \bibnamefont {Irwin}},\ }\href
  {\doibase 10.1103/PhysRevD.98.063004} {\bibfield  {journal} {\bibinfo
  {journal} {Phys. Rev. D}\ }\textbf {\bibinfo {volume} {98}},\ \bibinfo
  {pages} {063004} (\bibinfo {year} {2018})}\BibitemShut {NoStop}%
\bibitem [{\citenamefont {Lackey}\ and\ \citenamefont
  {Wade}(2015)}]{Lackey:2014fwa}%
  \BibitemOpen
  \bibfield  {author} {\bibinfo {author} {\bibfnamefont {B.~D.}\ \bibnamefont
  {Lackey}}\ and\ \bibinfo {author} {\bibfnamefont {L.}~\bibnamefont {Wade}},\
  }\href {\doibase 10.1103/PhysRevD.91.043002} {\bibfield  {journal} {\bibinfo
  {journal} {\prd}\ }\textbf {\bibinfo {volume} {91}},\ \bibinfo {pages}
  {043002} (\bibinfo {year} {2015})},\ \Eprint {http://arxiv.org/abs/1410.8866}
  {arXiv:1410.8866 [gr-qc]} \BibitemShut {NoStop}%
\bibitem [{\citenamefont {{Shapiro}}\ and\ \citenamefont
  {{Teukolsky}}(1983)}]{1983bhwd.book.....S}%
  \BibitemOpen
  \bibfield  {author} {\bibinfo {author} {\bibfnamefont {S.~L.}\ \bibnamefont
  {{Shapiro}}}\ and\ \bibinfo {author} {\bibfnamefont {S.~A.}\ \bibnamefont
  {{Teukolsky}}},\ }\href@noop {} {\emph {\bibinfo {title} {Black holes, white
  dwarfs, and neutron stars: They physics of compact objects.~New York,
  Wiley-Interscience, 1983, 663 p.}}}\ (\bibinfo {year} {1983})\BibitemShut
  {NoStop}%
\bibitem [{\citenamefont {Del~Pozzo}\ \emph {et~al.}(2013)\citenamefont
  {Del~Pozzo}, \citenamefont {Li}, \citenamefont {Agathos}, \citenamefont {Van
  Den~Broeck},\ and\ \citenamefont {Vitale}}]{PhysRevLett.111.071101}%
  \BibitemOpen
  \bibfield  {author} {\bibinfo {author} {\bibfnamefont {W.}~\bibnamefont
  {Del~Pozzo}}, \bibinfo {author} {\bibfnamefont {T.~G.~F.}\ \bibnamefont
  {Li}}, \bibinfo {author} {\bibfnamefont {M.}~\bibnamefont {Agathos}},
  \bibinfo {author} {\bibfnamefont {C.}~\bibnamefont {Van Den~Broeck}}, \ and\
  \bibinfo {author} {\bibfnamefont {S.}~\bibnamefont {Vitale}},\ }\href
  {\doibase 10.1103/PhysRevLett.111.071101} {\bibfield  {journal} {\bibinfo
  {journal} {Phys. Rev. Lett.}\ }\textbf {\bibinfo {volume} {111}},\ \bibinfo
  {pages} {071101} (\bibinfo {year} {2013})}\BibitemShut {NoStop}%
\bibitem [{\citenamefont {Wiringa}\ \emph {et~al.}(1988)\citenamefont
  {Wiringa}, \citenamefont {Fiks},\ and\ \citenamefont
  {Fabrocini}}]{PhysRevC.38.1010}%
  \BibitemOpen
  \bibfield  {author} {\bibinfo {author} {\bibfnamefont {R.~B.}\ \bibnamefont
  {Wiringa}}, \bibinfo {author} {\bibfnamefont {V.}~\bibnamefont {Fiks}}, \
  and\ \bibinfo {author} {\bibfnamefont {A.}~\bibnamefont {Fabrocini}},\ }\href
  {\doibase 10.1103/PhysRevC.38.1010} {\bibfield  {journal} {\bibinfo
  {journal} {Phys. Rev. C}\ }\textbf {\bibinfo {volume} {38}},\ \bibinfo
  {pages} {1010} (\bibinfo {year} {1988})}\BibitemShut {NoStop}%
\bibitem [{\citenamefont {{Akmal}}\ \emph {et~al.}(1998)\citenamefont
  {{Akmal}}, \citenamefont {{Pandharipande}},\ and\ \citenamefont
  {{Ravenhall}}}]{1998PhRvC..58.1804A}%
  \BibitemOpen
  \bibfield  {author} {\bibinfo {author} {\bibfnamefont {A.}~\bibnamefont
  {{Akmal}}}, \bibinfo {author} {\bibfnamefont {V.~R.}\ \bibnamefont
  {{Pandharipande}}}, \ and\ \bibinfo {author} {\bibfnamefont {D.~G.}\
  \bibnamefont {{Ravenhall}}},\ }\href {\doibase 10.1103/PhysRevC.58.1804}
  {\bibfield  {journal} {\bibinfo  {journal} {\prc}\ }\textbf {\bibinfo
  {volume} {58}},\ \bibinfo {pages} {1804} (\bibinfo {year} {1998})},\ \Eprint
  {http://arxiv.org/abs/nucl-th/9804027} {arXiv:nucl-th/9804027} \BibitemShut
  {NoStop}%
\bibitem [{\citenamefont {{M{\"u}ther}}\ \emph {et~al.}(1987)\citenamefont
  {{M{\"u}ther}}, \citenamefont {{Prakash}},\ and\ \citenamefont
  {{Ainsworth}}}]{1987PhLB..199..469M}%
  \BibitemOpen
  \bibfield  {author} {\bibinfo {author} {\bibfnamefont {H.}~\bibnamefont
  {{M{\"u}ther}}}, \bibinfo {author} {\bibfnamefont {M.}~\bibnamefont
  {{Prakash}}}, \ and\ \bibinfo {author} {\bibfnamefont {T.~L.}\ \bibnamefont
  {{Ainsworth}}},\ }\href {\doibase 10.1016/0370-2693(87)91611-X} {\bibfield
  {journal} {\bibinfo  {journal} {{Phys. Lett. B}}\ }\textbf {\bibinfo {volume}
  {199}},\ \bibinfo {pages} {469} (\bibinfo {year} {1987})}\BibitemShut
  {NoStop}%
\bibitem [{\citenamefont {{Lackey}}\ \emph {et~al.}(2006)\citenamefont
  {{Lackey}}, \citenamefont {{Nayyar}},\ and\ \citenamefont
  {{Owen}}}]{2006PhRvD..73b4021L}%
  \BibitemOpen
  \bibfield  {author} {\bibinfo {author} {\bibfnamefont {B.~D.}\ \bibnamefont
  {{Lackey}}}, \bibinfo {author} {\bibfnamefont {M.}~\bibnamefont {{Nayyar}}},
  \ and\ \bibinfo {author} {\bibfnamefont {B.~J.}\ \bibnamefont {{Owen}}},\
  }\href {\doibase 10.1103/PhysRevD.73.024021} {\bibfield  {journal} {\bibinfo
  {journal} {\prd}\ }\textbf {\bibinfo {volume} {73}},\ \bibinfo {eid} {024021}
  (\bibinfo {year} {2006})},\ \Eprint {http://arxiv.org/abs/astro-ph/0507312}
  {arXiv:astro-ph/0507312} \BibitemShut {NoStop}%
\bibitem [{\citenamefont {{M{\"u}ller}}\ and\ \citenamefont
  {{Serot}}(1996)}]{Mueller:1996pm}%
  \BibitemOpen
  \bibfield  {author} {\bibinfo {author} {\bibfnamefont {H.}~\bibnamefont
  {{M{\"u}ller}}}\ and\ \bibinfo {author} {\bibfnamefont {B.~D.}\ \bibnamefont
  {{Serot}}},\ }\href {\doibase 10.1016/0375-9474(96)00187-X} {\bibfield
  {journal} {\bibinfo  {journal} {Nucl. Phys.}\ }\textbf {\bibinfo {volume}
  {A606}},\ \bibinfo {pages} {508} (\bibinfo {year} {1996})},\ \Eprint
  {http://arxiv.org/abs/nucl-th/9603037} {arXiv:nucl-th/9603037 [nucl-th]}
  \BibitemShut {NoStop}%
\bibitem [{\citenamefont {{Steiner}}\ \emph {et~al.}(2010)\citenamefont
  {{Steiner}}, \citenamefont {{Lattimer}},\ and\ \citenamefont
  {{Brown}}}]{Steiner:2010}%
  \BibitemOpen
  \bibfield  {author} {\bibinfo {author} {\bibfnamefont {A.~W.}\ \bibnamefont
  {{Steiner}}}, \bibinfo {author} {\bibfnamefont {J.~M.}\ \bibnamefont
  {{Lattimer}}}, \ and\ \bibinfo {author} {\bibfnamefont {E.~F.}\ \bibnamefont
  {{Brown}}},\ }\href {\doibase 10.1088/0004-637X/722/1/33} {\bibfield
  {journal} {\bibinfo  {journal} {\apj}\ }\textbf {\bibinfo {volume} {722}},\
  \bibinfo {pages} {33} (\bibinfo {year} {2010})},\ \Eprint
  {http://arxiv.org/abs/1005.0811} {arXiv:1005.0811 [astro-ph.HE]} \BibitemShut
  {NoStop}%
\bibitem [{\citenamefont {Nattila}\ \emph {et~al.}(2017)\citenamefont
  {Nattila}, \citenamefont {Miller}, \citenamefont {Steiner}, \citenamefont
  {Kajava}, \citenamefont {Suleimanov},\ and\ \citenamefont
  {Poutanen}}]{Nattila:2017wtj}%
  \BibitemOpen
  \bibfield  {author} {\bibinfo {author} {\bibfnamefont {J.}~\bibnamefont
  {Nattila}}, \bibinfo {author} {\bibfnamefont {M.~C.}\ \bibnamefont {Miller}},
  \bibinfo {author} {\bibfnamefont {A.~W.}\ \bibnamefont {Steiner}}, \bibinfo
  {author} {\bibfnamefont {J.~J.~E.}\ \bibnamefont {Kajava}}, \bibinfo {author}
  {\bibfnamefont {V.~F.}\ \bibnamefont {Suleimanov}}, \ and\ \bibinfo {author}
  {\bibfnamefont {J.}~\bibnamefont {Poutanen}},\ }\href {\doibase
  10.1051/0004-6361/201731082} {\bibfield  {journal} {\bibinfo  {journal}
  {Astron. Astrophys.}\ }\textbf {\bibinfo {volume} {608}},\ \bibinfo {pages}
  {A31} (\bibinfo {year} {2017})},\ \Eprint {http://arxiv.org/abs/1709.09120}
  {arXiv:1709.09120 [astro-ph.HE]} \BibitemShut {NoStop}%
\bibitem [{\citenamefont {Shibata}\ \emph {et~al.}(2017)\citenamefont
  {Shibata}, \citenamefont {Fujibayashi}, \citenamefont {Hotokezaka},
  \citenamefont {Kiuchi}, \citenamefont {Kyutoku}, \citenamefont {Sekiguchi},\
  and\ \citenamefont {Tanaka}}]{Shibata:2017xdx}%
  \BibitemOpen
  \bibfield  {author} {\bibinfo {author} {\bibfnamefont {M.}~\bibnamefont
  {Shibata}}, \bibinfo {author} {\bibfnamefont {S.}~\bibnamefont
  {Fujibayashi}}, \bibinfo {author} {\bibfnamefont {K.}~\bibnamefont
  {Hotokezaka}}, \bibinfo {author} {\bibfnamefont {K.}~\bibnamefont {Kiuchi}},
  \bibinfo {author} {\bibfnamefont {K.}~\bibnamefont {Kyutoku}}, \bibinfo
  {author} {\bibfnamefont {Y.}~\bibnamefont {Sekiguchi}}, \ and\ \bibinfo
  {author} {\bibfnamefont {M.}~\bibnamefont {Tanaka}},\ }\href {\doibase
  10.1103/PhysRevD.96.123012} {\bibfield  {journal} {\bibinfo  {journal} {Phys.
  Rev.}\ }\textbf {\bibinfo {volume} {D96}},\ \bibinfo {pages} {123012}
  (\bibinfo {year} {2017})},\ \Eprint {http://arxiv.org/abs/1710.07579}
  {arXiv:1710.07579 [astro-ph.HE]} \BibitemShut {NoStop}%
\bibitem [{\citenamefont {Radice}\ \emph {et~al.}(2018)\citenamefont {Radice},
  \citenamefont {Perego}, \citenamefont {Zappa},\ and\ \citenamefont
  {Bernuzzi}}]{Radice:2017lry}%
  \BibitemOpen
  \bibfield  {author} {\bibinfo {author} {\bibfnamefont {D.}~\bibnamefont
  {Radice}}, \bibinfo {author} {\bibfnamefont {A.}~\bibnamefont {Perego}},
  \bibinfo {author} {\bibfnamefont {F.}~\bibnamefont {Zappa}}, \ and\ \bibinfo
  {author} {\bibfnamefont {S.}~\bibnamefont {Bernuzzi}},\ }\href {\doibase
  10.3847/2041-8213/aaa402} {\bibfield  {journal} {\bibinfo  {journal}
  {Astrophys. J.}\ }\textbf {\bibinfo {volume} {852}},\ \bibinfo {pages} {L29}
  (\bibinfo {year} {2018})},\ \Eprint {http://arxiv.org/abs/1711.03647}
  {arXiv:1711.03647 [astro-ph.HE]} \BibitemShut {NoStop}%
\bibitem [{\citenamefont {Coughlin}\ \emph {et~al.}(2018)\citenamefont
  {Coughlin} \emph {et~al.}}]{Coughlin:2018miv}%
  \BibitemOpen
  \bibfield  {author} {\bibinfo {author} {\bibfnamefont {M.~W.}\ \bibnamefont
  {Coughlin}} \emph {et~al.},\ }\href@noop {} {\  (\bibinfo {year} {2018})},\
  \Eprint {http://arxiv.org/abs/1805.09371} {arXiv:1805.09371 [astro-ph.HE]}
  \BibitemShut {NoStop}%
\bibitem [{\citenamefont {{Gravitational Wave Open Science Center
  (GWOSC)}}(2017)}]{GWOSC:GW170817}%
  \BibitemOpen
  \bibfield  {author} {\bibinfo {author} {\bibnamefont {{Gravitational Wave
  Open Science Center (GWOSC)}}},\ }\href@noop {} {}\bibinfo {howpublished}
  {\url{https://doi.org/10.7935/K5B8566F}} (\bibinfo {year} {2017})\BibitemShut
  {NoStop}%
\end{thebibliography}%

\iftoggle{endauthorlist}{
  %
  %
  \let\author\myauthor
  \let\affiliation\myaffiliation
  \let\maketitle\mymaketitle
  \title{Authors}
  \pacs{}
  
 \newpage
  \maketitle
}

\end{document}